\documentclass[structabstract]{aa}  
\usepackage{graphicx}
\usepackage{txfonts}
\usepackage{color}
\usepackage{enumerate}
\usepackage{lscape}
\begin{document}
   \title{Overview of semi-sinusoidal stellar variability \\ with the CoRoT satellite
\thanks{
The CoRoT space mission was developed and is operated by the French
space agency CNES, with the participation of ESA's RSSD and Science Programmes,
Austria, Belgium, Brazil, Germany, and Spain.}}

\author{J. R. De Medeiros \inst{1}
\and {C. E. Ferreira Lopes \inst{1,2,3}}
\and I. C. Le\~ao \inst{1}
\and B. L. Canto Martins \inst{1}
\and M. Catelan \inst{2,3}
\and A. Baglin \inst{4}
\and S. Vieira \inst{1}
\and J.~P.~Bravo \inst{1}
\and C. Cort\'es \inst{5,1}
\and D. B. de Freitas \inst{1}
\and E. Janot-Pacheco \inst{6}
\and S.~C.~Maciel \inst{7,1}
\and C. H. F. Melo \inst{8}
\and Y. Osorio \inst{9}
\and G.~F.~Porto~de~Mello \inst{10}
\and A. Valio \inst{11}
}

\authorrunning{J. R. De Medeiros et al.}
\titlerunning{Semi-sinusiodal variability with the CoRoT satellite}

   \institute{Departamento de F\'isica, Universidade Federal do Rio Grande do Norte, Natal, RN, 59072-970 Brazil\\
              \email{renan@dfte.ufrn.br}
   \and Departamento de Astronom\'ia y Astrof\'isica, Pontificia Universidad Cat\'olica de Chile, Av. Vicu\~na Mackenna 4860, 782-0436 Macul, Santiago, Chile
   \and The Milky Way Millennium Nucleus, Av. Vicu\~{n}a Mackenna 4860, 782-0436 Macul, Santiago, Chile
   \and LESIA, UMR 8109 CNRS, Observatoire de Paris, UVSQ, Universit\'e Paris-Diderot, 5 place J. Janssen, 92195 Meudon, France
   \and Departamento de F\'isica, Facultad de Ciencias B\'asicas, Universidad Metropolitana de la Educacion, Av. Jos\'e Pedro Alessandri 774, 7760197, \~Nu\~noa, Santiago, Chile
   \and Universidade de S\~ao Paulo/IAG-USP, rua do Mat\~ao, 1226, Cidade Universit\'aria, S\~ao Paulo, SP, 05508-900 Brazil
   \and IFPB - Instituto Federal de Educa\c{c}\~ao, Ci\^encia e Tecnologia da Para\'iba Av. Primeiro de Maio, 720,  Jo\~ao Pessoa - PB, 58015-430, Brazil
   \and European Southern Observatory, Casilla 19001, Santiago, Chile
   \and Department of Physics and Astronomy Uppsala University, Box 516, 751 20, Uppsala, Sweden
   \and Universidade Federal do Rio de Janeiro, Observat\'orio do Valongo, Ladeira do Pedro Antonio, 43 Rio de Janeiro, 20080-090 Brazil
   \and Center for Radio Astronomy and Astrophysics Mackenzie, Universidade Presbiteriana Mackenzie, Rua da Consola\c{c}\~ao, 896 S\~ao Paulo, SP, Brazil }

   \date{Received Month Day, Year; accepted Month Day, Year}

 
  \abstract
   {To date, the CoRoT space mission has produced more than 124,471 light 
curves. Classifying these curves in terms of unambiguous variability 
behavior is mandatory for obtaining an unbiased statistical view on their 
controlling root-causes.}
   {The present study provides an overview of 
semi-sinusoidal light curves observed by the CoRoT exo--field CCDs.}
   {We selected a sample of 4,206 light curves presenting well-defined semi-sinusoidal signatures.
The variability periods were computed based on Lomb-Scargle periodograms, harmonic
fits, and visual inspection.}
   {Color-period diagrams for the present sample show
the trend of an increase of the variability periods as long as the stars evolve.
This evolutionary behavior is also noticed when comparing the
period distribution in the Galactic center and anti-center directions.
These aspects indicate a compatibility
with stellar rotation, although more information is needed to
confirm their root-causes.
Considering this possibility, we
identified a subset of three Sun-like candidates by their photometric period.
Finally, the variability period versus color 
diagram behavior was found to be highly dependent on the reddening correction.}
   {}


   \maketitle
%

\section{Introduction}

The CoRoT space mission has been operational for more than three years (Baglin et al. 2009). Its main science goals are asteroseismology and the search for exoplanets based 
on transit detection. Thanks to this space mission, a unique set of light curves (LCs) is now available for about 140,000 stars, with excellent time-sampling and unprecedented 
photometric precision. The photometric data obtained are a rich source for different astrophysical studies. 
For instance, the luminosity of stars can vary for a number of reasons, including gravitational deformation and eclipses due to binarity, as well as surface oscillations and rotation resulting from star spots.
The variability induced by each phenomenon has a characteristic range of time scales and amplitudes.
Open questions of high interest in this area include the behavior of stellar rotation periods, differential rotation as a function of latitude, distribution of spot areal coverage, the spot distributions in longitude and latitude on different stars, the presence and distribution of active longitudes, the timescale for evolution of different-sized spots, spot contrasts, and the evolutionary behavior of all these as a function of stellar mass, age, and metallicity. 

An initial overview of stellar variability in CoRoT data was described in Debosscher et al.~(\cite{deb07}, \cite{deb09}), based on automated supervised classification methods for variable stars. The authors described a significant fraction of (quasi-) monoperiodic variables with low amplitude in the first four measured fields of the CoRoT exoplanet program. The majority are most likely rotationally modulated variables, with some low-amplitude Cepheids. Nevertheless, as reported
by these authors, automatic procedures offer variability classification that is sensitive to different artifacts. Therefore, misclassification always occurs and its incidence depends significantly on the variability class considered. This is particularly true for variability caused by star spots, which are highly dynamic, and reflects the effects of surface rotation, differential rotation, spot lifetime, and transient phenomena such as flares, primarily for lower mass main-sequence stars (e.g.,~Lanza et al.~\cite{lan07}; Hartman et al.~\cite{har09}).
For example, star spots and photometric rotational modulation have long been studied using photometry or spectroscopy (Strassmeier \cite{str09}; Hartman et al.~\cite{har10}; Meibom et al.~\cite{mei09}, \cite{mei11}; Irwin et al.~\cite{irw11}). However, ground-based observations result in a number of time gaps,
and the coverage area of spots on star surfaces must be several times larger than that on the Sun to achieve a robust signal. 
More recently, Affer et al.~(\cite{aff12}) presented rotation period measurements for 1727 CoRoT field stars and claimed to have identified a sample of young stars ($<$ 600 Myr).

The present study describes the first results of our effort to determine variability periods from the LCs of stars in the exoplanet fields observed with CoRoT. 
Of the 124,471 LCs produced to date from all CoRoT observing runs, we selected those displaying semi--sinusoidal variations with particular patterns to be described in Sect.~\ref{finalselec}.
The first portion of this paper is devoted to describing the stellar sample, the observations, and the procedure for determining the variability period. Our results are presented together with global analyses based on available stellar parameters, such as color index, luminosity classes, and spectral types. Here, we describe our unprecedented contribution to the treatment of CoRoT LCs, which was the correction of reddening effects on the stellar colors. In fact, CoRoT was designed to observe stellar samples in the Galactic center and anti-center directions. The exo-fields typically observe relatively faint stars with V~$\sim$ 11 to 16 mag. This suggests that CoRoT targets may be subject to considerable interstellar extinction effects. The reddening effect on CoRoT targets has not been explored yet. Careful analysis of these effects is mandatory, for example, to minimize possible bias on color-period behavior of variability distributions, as well as the location of stars in color-period diagrams. In particular, evolutionary scenarios of stellar variability parameters can certainly be better understood when a reddening correction is performed. Finally, our main conclusions are presented, in addition to the primary goals for future studies.

%
\section{Working sample, observations, and data analysis}
\label{secobs}

Raw LCs are collected by the CoRoT satellite as N0 data and after they are
processed on the ground by the CoRoT pipeline (Samadi et al.~\cite{sam07}) in two levels.
In the first level some electronic, background, and jitter effects are corrected,
and data taken during on the South Atlantic Anomaly (SAA) passage are flagged, producing the N1 data.
Subsequent treatments are proceessed in the second level and include sampling combination, calculation of heliocentric date,
and flagging of hot pixels. The results are the N2
data, which are provided to the public for science analysis.

\begin{table}
	\centering
		\begin{tabular}{c|c|c}
		 \hline
      CoRoT Run & Total LCs & Total time span (days) \\
     \hline
     IRa01	&	9880	&	54-57	\\
     LRa01	&	11408	&	131	\\
     LRa02	&	11408	&	111-114	\\
     LRa03	&	5289	&	148	\\
     LRc01	&	11407	&	142-152	\\
     LRc02	&	11408	&	144	\\
     LRc03	&	5661	&	89	\\
     LRc04	&	5716	&	84	\\
     LRc05	&	5683	&	87	\\
     LRc06	&	5683	&	77	\\
     SRa01	&	8150	&	23	\\
     SRa02	&	10265	&	31	\\
     SRa03	&	4130	&	24	\\
     SRc01	&	6975	&	25	\\
     SRc02	&	11408	&	20	\\	
    \hline
		\end{tabular}
	\caption{Basic properties of the dataset analyzed by CoRoT, indicating the number of LCs in each observing run and the respective total span time, totaling 124,471 LCs. The lower-case ``a'' means the Galactic anti-center and ``c'' means the Galactic center direction.}
	\label{tab_sample}
\end{table}

For this investigation we selected the calibrated LCs measured with CoRoT exoplanet CCDs during 3 years of operation, with stars exhibiting visual magnitudes ranging from about 12 to 16. Time sampling for the LCs is 32 s, but for most data an average is calculated over 16 such measurements, resulting in an effective time resolution of 512 s. For a fraction of the LCs (or in some cases, parts of them), the original 32 s sampling was retained. These LCs correspond to high-priority targets measured in oversampling mode, totaling approximately 124,471 CoRoT LCs from the Initial Run (IR), Long Runs (LR), and Short Runs (SR), with a time window of between 20 and 157 days. Additional basic properties of the data are listed in Table~\ref{tab_sample}.

\subsection{Data treatment}

The CoRoT pipeline provides N2 LCs corrected for several effects,
but still with some problems that need additional treatment before the science analysis.
In particular, CoRoT N2 LCs may have jumps (discontinuities) produced by hot pixels,
long-term trends produced by CCD temperature variations, and outliers.
There is no standard method for those post-treatments and different works have their methods
according to their objectives (e.g.,~Renner et al.~\cite{ren08};  Basri et al.~\cite{bas11}; Affer et al.~\cite{aff12}).
We describe below our procedure performed for data treatment, selection, and analysis.
Our procedure is basically a set of steps and rules that were mainly performed manually.
In particular, a simplified automatic version of this procedure was run in the beginning
for a preliminary sample selection,
then the selected sub-sample was re-treated and re-analyzed manually step by step.

We considered the LCs in normalized flux units, namely $F$,
dividing each LC by its whole flux average.
We then defined the noise level $\sigma$ of each LC
as being
simply a high-frequency contribution obtained from the standard deviation of the difference between the nearest-neighbor flux measurements, which yields

\begin{equation}
\label{eq:noise}
  \sigma = \sqrt{\frac{1}{N} \sum_{i=1}^N{(F_i - F_{i-1})^2}},
\end{equation}
where $F_i$ is the flux value corresponding to the observing time $t_i$,
and we considered $F_0 \equiv F_N$.
For a homogeneous noise calculation the LCs were resampled to a bin of 864~s (0.01~days)\footnote{This bin also saved computation time and did not affect the frequencies considered in the data analysis~(Sect.~\ref{subanalysis}).}.
The next steps were the data treatment,
where we first performed a jump correction,
followed by a long-term detrend
and, finally, the removal of outliers.
These steps are explained below.

\subsubsection{Jump correction}
\label{jumpcorr}

\begin{figure}
\begin{center}
\includegraphics[width=\columnwidth,height=4cm]{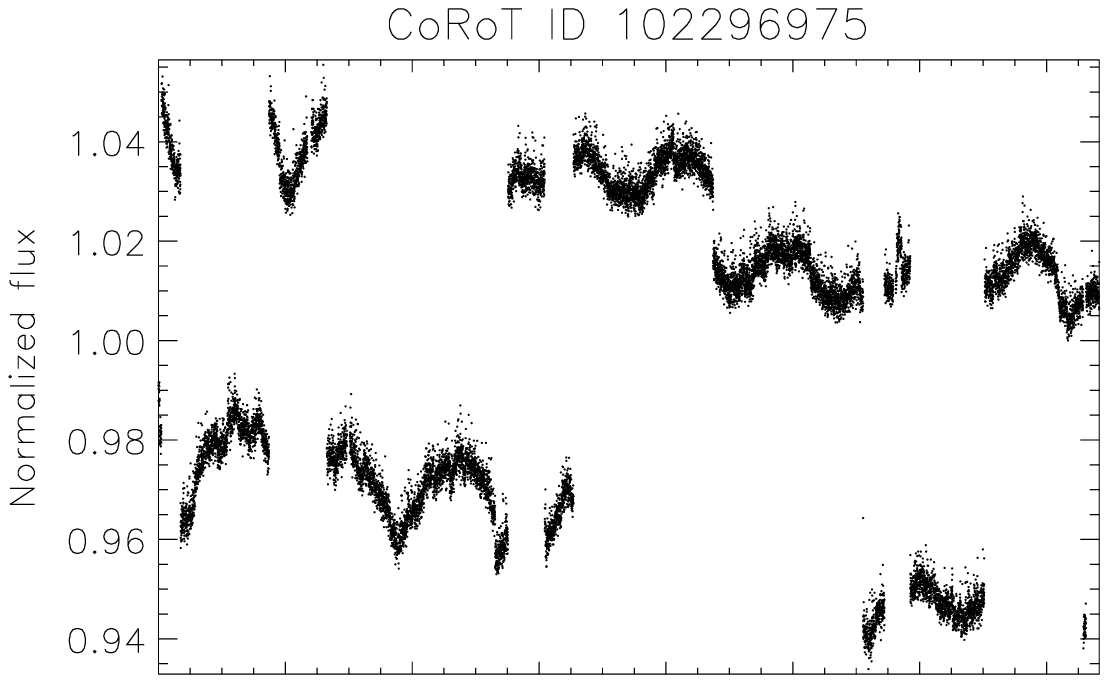}
\includegraphics[width=\columnwidth,height=4cm]{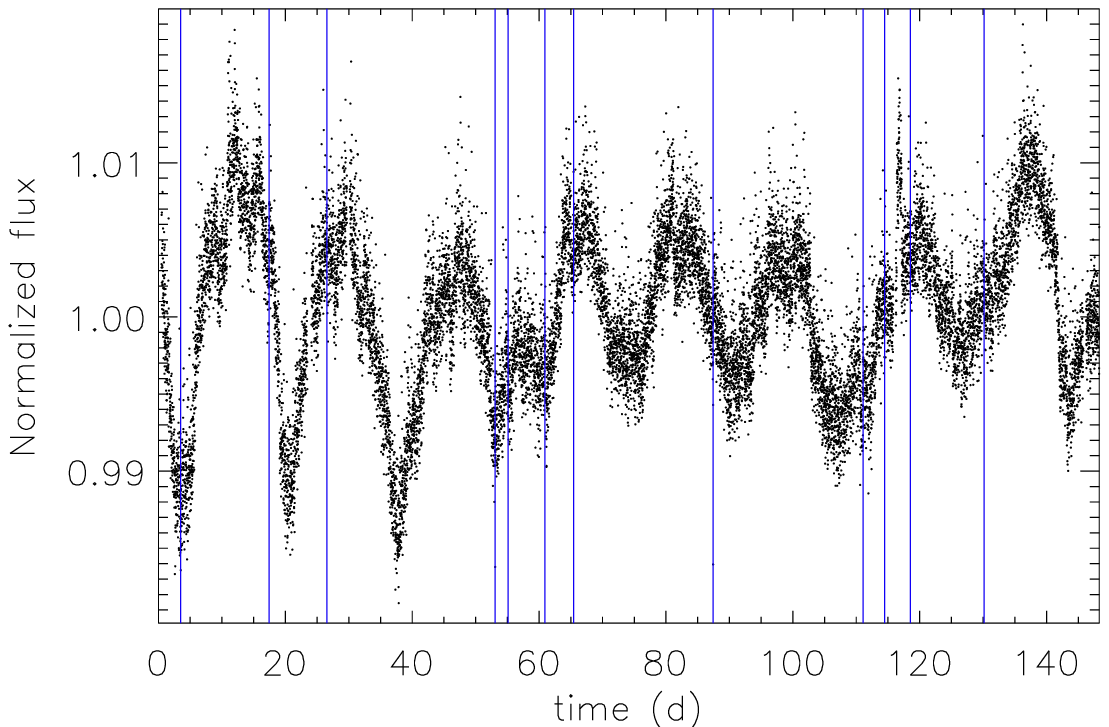}
\caption{
Example of a LC with several jumps. Top panel: original LC.
Bottom panel: LC after the jump correction described in Sec.~\ref{jumpcorr},
where the vertical lines indicate the corrected jumps.}
\label{figjumps}
\end{center}
\end{figure}

The CoRoT pipeline corrects for some jumps, but some still remain in the N2 LCs.
These discontinuities may be caused by a sudden change of the mean LC level in a single (or few) time step(s).
Therefore a LC may have several jumps.
There is no unique method for detecting and correcting these discontinuities.
Different algorithms were proposed to correct for them, usually in the search of planetary transits (eg.~Mislis et al.~\cite{mis10}).
For some other types of stellar variabilities,
these corrections may be more challenging because we must keep the information of
smoother and more irregular variations than those caused by transits.
We developed a method for detecting and correcting jumps
that combined with visual inspection can be used in the correction of most cases.
Fig.~\ref{figjumps} shows an example of an LC with several jumps that hide any physical information.
After the jump correction with this method, here applied automatically, the variability signature becomes noticable.
This method worked well automatically in this example and in several cases,
although it produces miscorrections in some cases.
Nevertheless, we checked in practice that for the large majority of LCs any mismatch produces a lower power in the periodogram than the main variation (e.g., in Fig.\ref{figjumps}) and does not strongly affect the determination of the variability parameters.
After a first automatic filtering,
a manual correction was applied by testing different levels of corrections when analyzing the LCs, periodograms, and phase diagrams.
Doubtful cases were simply rejected.

To determine if there was a jump within a time interval
$t_{i-1}$ and $t_i$, we considered a box of duration $\Delta t$ both
to the left (previous) and to the right (after) of the time interval.
First the mean flux was estimated within each box separately. When the
difference between the left and right flux averages, $\Delta F$, was
greater than a defined threshold,  $\Delta F_J$, then a discontinuity
was assumed to occur from $t_{i-1}$ to $t_i$.  To avoid correcting for
false jumps caused by a very steep flux variation within the LC,
a linear fit was performed independently on the data contained in both
boxes of duration $\Delta t$. The higher of the two angular
coefficients of the fits was assumed as $\left|\delta F / \delta t
\right|_{\rm max}$, the estimated rate of flux variation between the
boxes. The jump threshold, $\Delta F_J$, was defined as

\begin{equation}
\label{eq:J}
  \Delta F_J = a \sigma + b \left|\delta F / \delta t \right|_{\rm max} (t_i-t_{i-1}),
\end{equation}
where $\sigma$ is the noise level, whereas $a$ and $b$ are constants.
To correct for the discontinuity when $\Delta F > \Delta F_J$,
we considered two boxes of a short duration $\Delta t_s < \Delta t$
to the left and right of the detected jump.
The flux levels of the left and the right box
were adjusted so as to make the box averages equal.
Our experience showed that the box
durations of $\Delta t = 1$~day and $\Delta t_s = 0.1$~day and a threshold level with $a \simeq 4$
and $b \simeq 2$ were capable of detecting and correcting most jumps.

\subsubsection{Long-term trends and outliers}

After the jump correction, we minimized long-term trends by dividing the LC
by a third-order polynomial fit,
as performed in previous works (e.g.,~Basri et al.~\cite{bas11}; Affer et al.~\cite{aff12}).
Finally, we removed some outliers with flux values that typically differed by more than about five times
the standard deviation of a LC.
From this point on, a LC was considered to be fully treated and its analysis could be performed.

%
\subsection{Light curve analysis and selection}
\label{subanalysis}



To properly analyze the LCs and select their parameters, we developed simple noise-free LC models from harmonic fits similar to those described in Debosscher et al.~(\cite{deb07}).
For each LC, the Lomb-Scargle periodogram (Lomb~\cite{lom76}; Scargle~\cite{sca82}) was computed for periods with a false alarm probability FAP $< 0.01$  (significance level $>$~99\%).
The highest periodogram peak, named frequency $f_1$ or period $P_1$,
was refined to a near frequency with the highest ratio of amplitude (calculated from a harmonic fit of the phase diagram; see below and Sect.~\ref{perdet}) to the minimum dispersion. (computed from Eq.~(2) given in Dworetsky~\cite{dwo83}) of the phase diagram\footnote{This adjustment reduces numerical errors such as those originating from the periodogram resolution and from the LC time window.}.
Next, the refined frequency $f_1$
was used to calculate a harmonic fit with four harmonics.
The fit was computed from a non-linear least-squares minimization using the Levenberg-Marquardt method (Levenberg~\cite{lev44}; Marquardt~\cite{mar63}).
It was used to estimate a preliminary variability period and mean amplitude together with their errors.
The final period was not necessarily $P_1$, as we explain in Sect.~\ref{perdet}.
Based on this fit, a mean signal--to--noise ratio (S/N) of the LC was estimated as

\begin{equation}
S/N = \frac{A({\rm mag})}{\sigma({\rm mag})},
\label{eq_sn}
\end{equation}
where $A$(mag) is the mean variability amplitude in units of magnitude
and $\sigma$(mag) is the mean LC noise defined in Eq.~(\ref{eq:noise}) and converted to magnitude.

For the LC, the fit was then subtracted from the time series (prewhitening) and a new Lomb-Scargle periodogram was computed. The same procedure was repeated in ten iterations, finding ten independent frequencies, each with a harmonic fit of four harmonics.
These ten independent frequencies were then used to obtain a harmonic best fit with the original (trend-subtracted) time series as follows:

\begin{equation}
y(t) = \sum_{i = 1}^{10}\sum_{j = 1}^{4}\left[ a_{ij}\sin\left(2\pi f_{i} j t \right) + b_{ij}\cos\left(2\pi f_{i} j t \right) \right] + b_0,
\label{eq_best_harm}
\end{equation}
where $a_{ij}$ and $b_{ij}$ are Fourier coefficients, $t$ is the time and $b_0$ is the background level.
The choice of four harmonics and ten iterations is based on a compromise between a good fit and computation time.
This harmonic fit was used as a model to
analyze some temporal variations on the amplitude and other aspects.
These aspects are specifically criteria (iv) to (vi) of Sect.~\ref{semisinus}.

\subsubsection{Selection by S/N}
\label{snrsel}

\begin{figure}
\begin{center}
\includegraphics[width=\columnwidth,height=4cm]{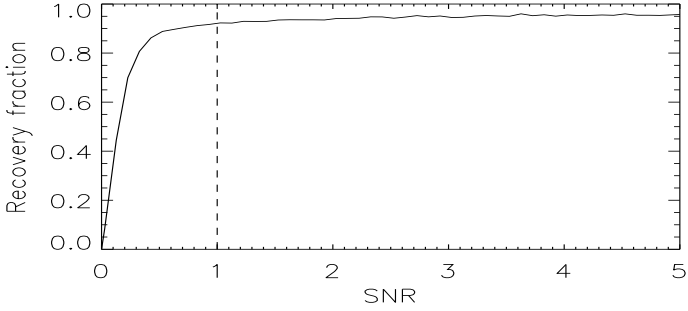}
\caption{
Recovery fraction for period determination as a function of S/N.
The vertical dashed line indicates the S/N ($\sim$1.0) above which the recovery fraction is close to the maximum.
}
\label{figsnr}
\end{center}
\end{figure}

In the simplified automatic procedure, the methods described above were run in the beginning to estimate
the mean S/N of the LCs and select a first subsample.
To define a proper cut-off value for the S/N, 
we determined the reliability of a variability period as a function of the S/N
by testing several simulations of semi-sinusoidal variabilities (300,000 simulations).
These simulations were random pieces of actual CoRoT LCs with variabilities showing more than five cycles and S/N~$>$~5
(that were assumed to have a good period determination, namely ``true'' periods),
extracted from our own sample.
For each simulated LC, the high-frequency signal -- assumed to be the noise -- was amplified or reduced by a random factor
to change the S/N.
Next, the Lomb-Scargle periodogram was computed and the most significant peak was compared with the ``true'' period from the simulation.
We then counted how many times the simulation periods were correctly recovered.
The recovery fraction as a function of the S/N is shown in figure~\ref{figsnr}.

In practice, figure~\ref{figsnr} indicates the likelihood of determining the correct variability period
as a function of the S/N of an LC.
This probability rises with increasing S/N and usually does not reach 100\% even with several cycles.
The highest probablity occurs for S/N~$\gtrsim$~1.0, thus,
only LCs with S/N above this cut-off value were selected.
The reason why the recovery fraction does not reach 100\% is explained below.

\subsubsection{Period determination}
\label{perdet}

\begin{figure}
\begin{center}
\includegraphics[width=\columnwidth,height=4cm]{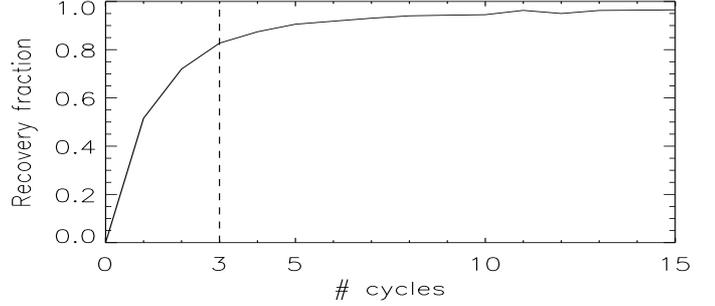}
\caption{
Recovery fraction for period determination as a function of the number of cycles.
The vertical dashed line indicates the number of cycles ($\sim$3) above which the recovery fraction is greater than $\sim$80\%.
}
\label{figcycles}
\end{center}
\end{figure}

An important problem in the variability period determination
is the fact that the observed period may be an alias or harmonic of the actual period (e.g.,~Hartman et al.~\cite{har10}).
Aliases are seen as several discrete peaks in the periodogram of a LC.
In some cases, selecting the correct period among harmonics or aliases may be ambiguous,
and choosing the correct peak is generally a difficult task.
Aliases may be avoided when the time window of the LC is long enough to present several cycles of the variability under analysis.
Indeed, the greater the number of observed cycles, the better the determination of variability period.

Therefore, we determined the reliability of a variability period as a function of the number of cycles observed
by testing several simulations of semi-sinusoidal variabilities (300,000 simulations).
Similarly to the S/N analysis (Sect.~\ref{snrsel}),
random pieces of actual CoRoT LCs were taken, in this case with variabilities showing more than ten cycles
(to consider the best period determination, namely ``true'' periods) in our sample with an S/N~$>$~1.0.
The recovery fraction as a function of the number of cycles is shown in figure~\ref{figcycles}.
This probability, $\mathbf{\rho}$, rises with increasing number of cycles and does not reach 100\%, as in the S/N analysis.
Based on this figure, variabilities were divided into two groups: a higher confidence group, with more than three observed cycles in their LCs and $\rho \gtrsim$~80\%, and a lower confidence group, exhibiting less than three observed cycles in their LCs.

\begin{figure}
\begin{center}
\includegraphics[width=\columnwidth,height=4cm]{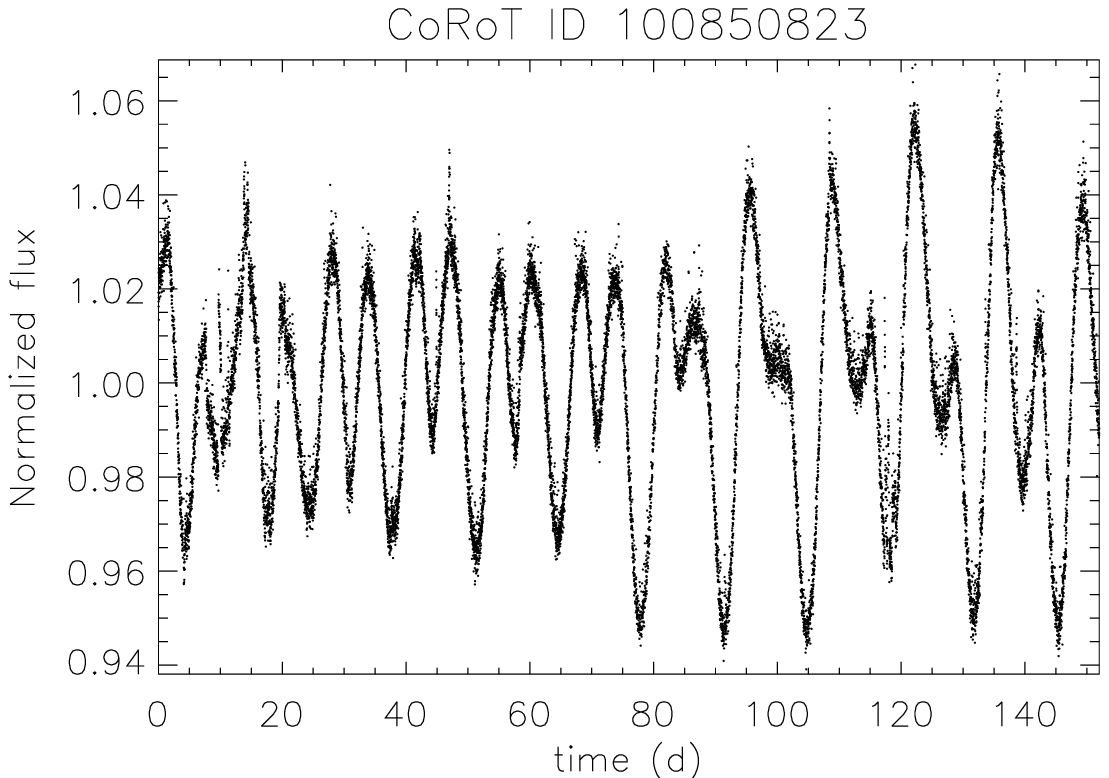}
\includegraphics[width=\columnwidth,height=4cm]{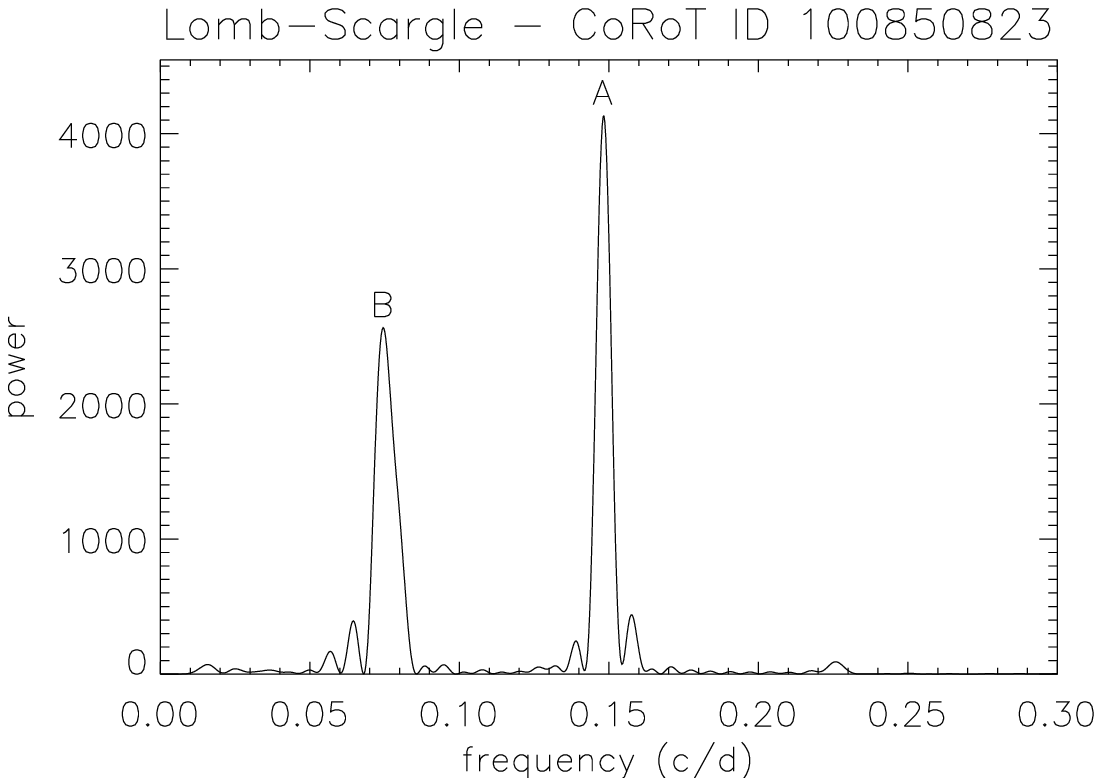}
\includegraphics[width=\columnwidth,height=4cm]{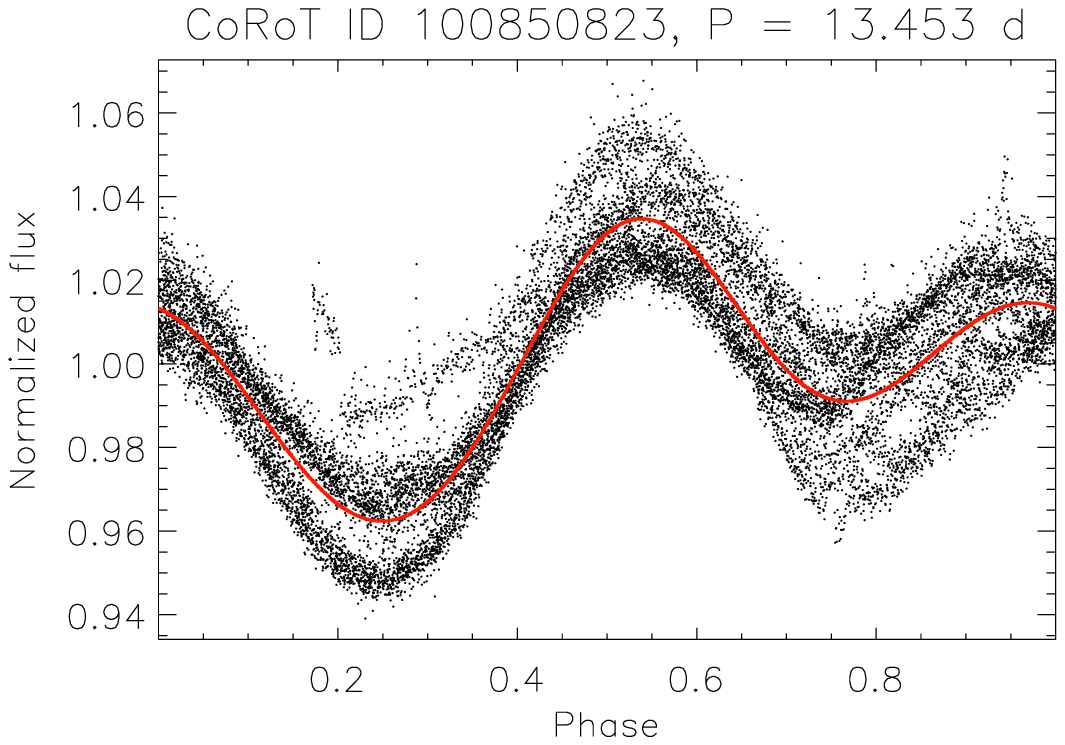}
\caption{
Example of a LC interpreted as having two subcycles per cycle.
Top panel: original LC. Medium panel: Lomb-Scargle periodogram showing the main
peak (A) with a period of 6.75 days, and a second peak (B) with a period of 13.4 days.
Bottom panel: phase diagram for a period of 13.453 days (adjusted to the highest amplitude and lowest dispersion),
showing one full cycle with two subcycles and a harmonic fit depicted by the solid line.
}
\label{figalias}
\end{center}
\end{figure}

\begin{figure}[t]
\begin{center}
\includegraphics[width=4.3cm]{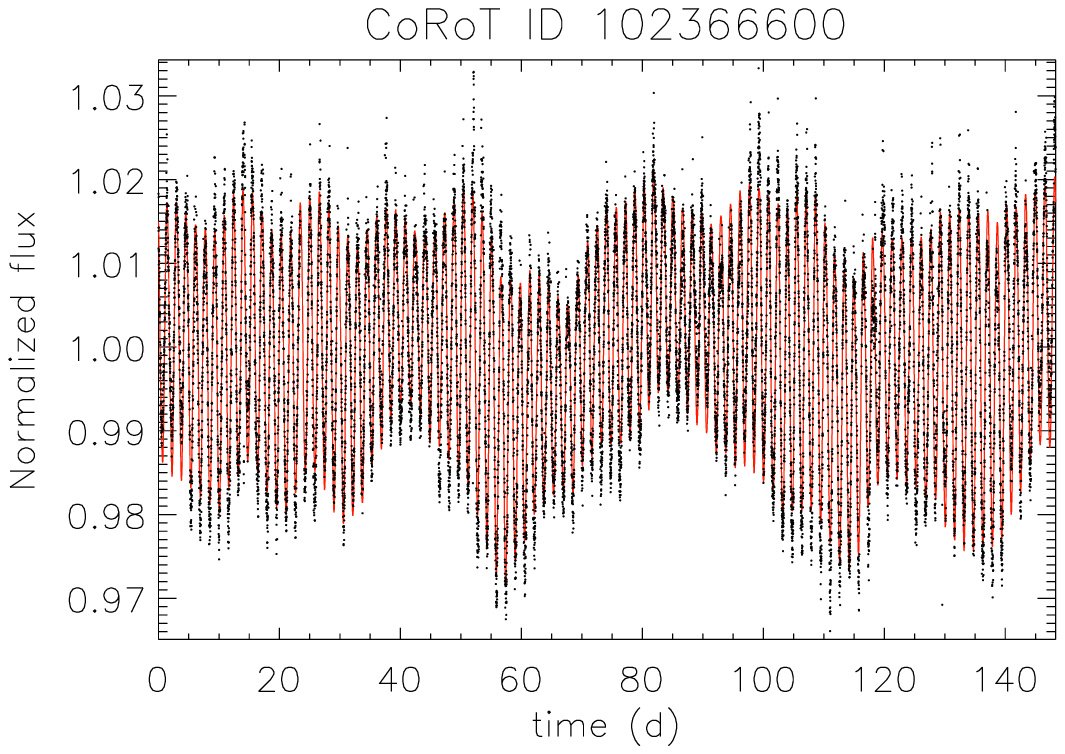}
\includegraphics[width=4.3cm]{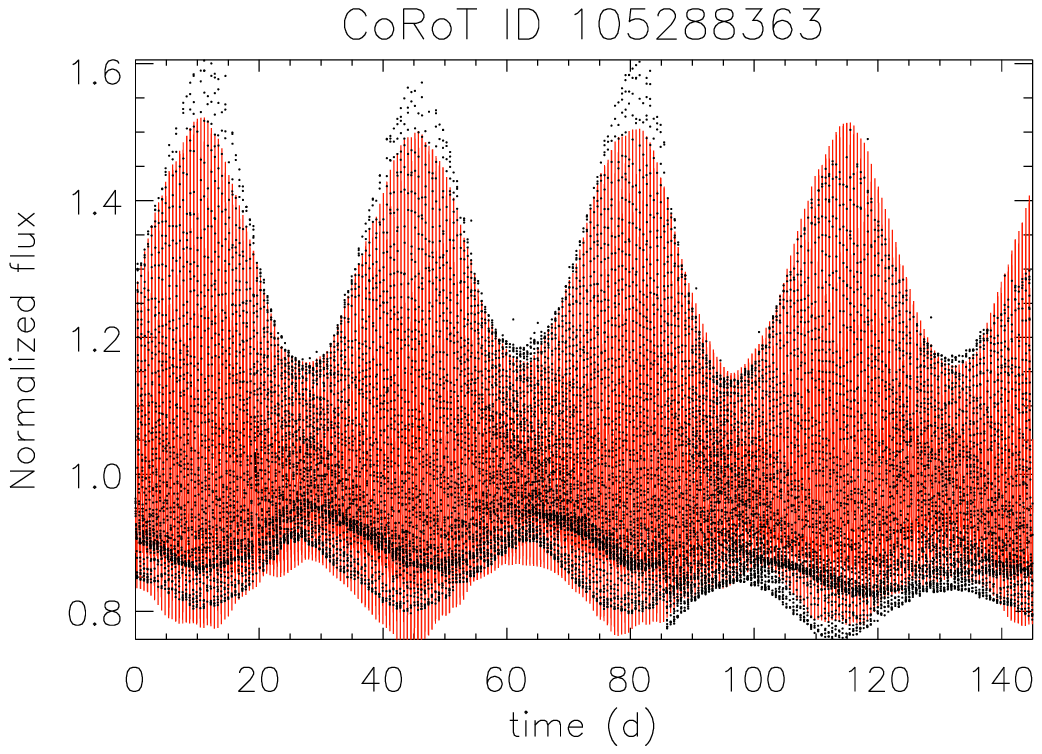} \\
\includegraphics[width=4.3cm]{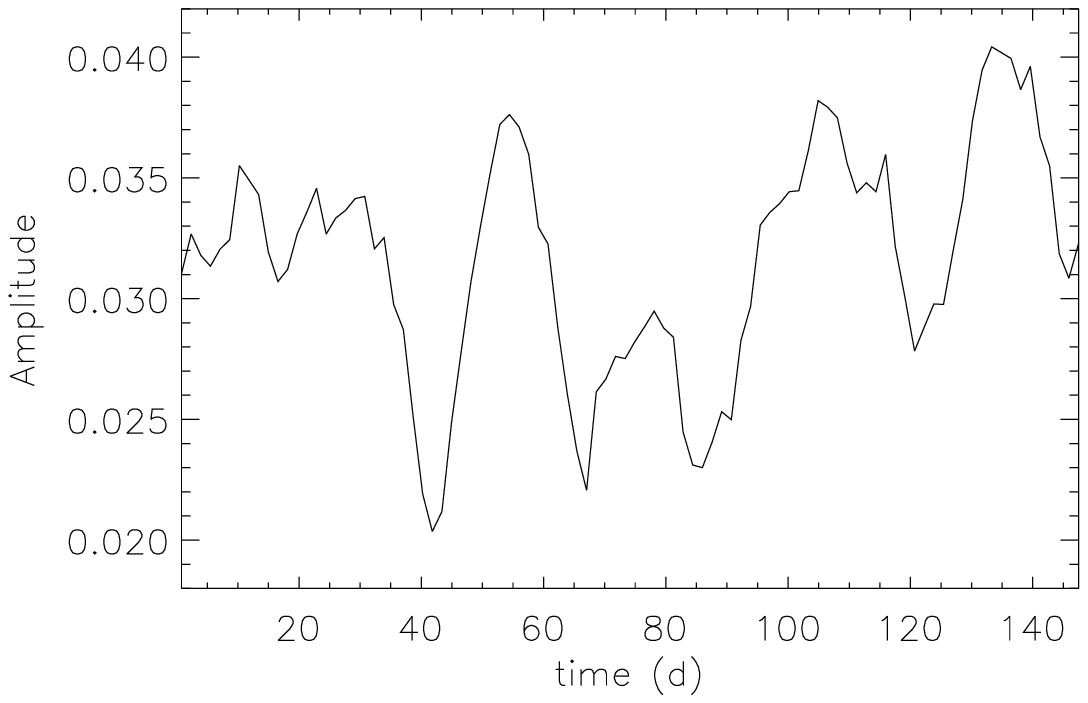}
\includegraphics[width=4.3cm]{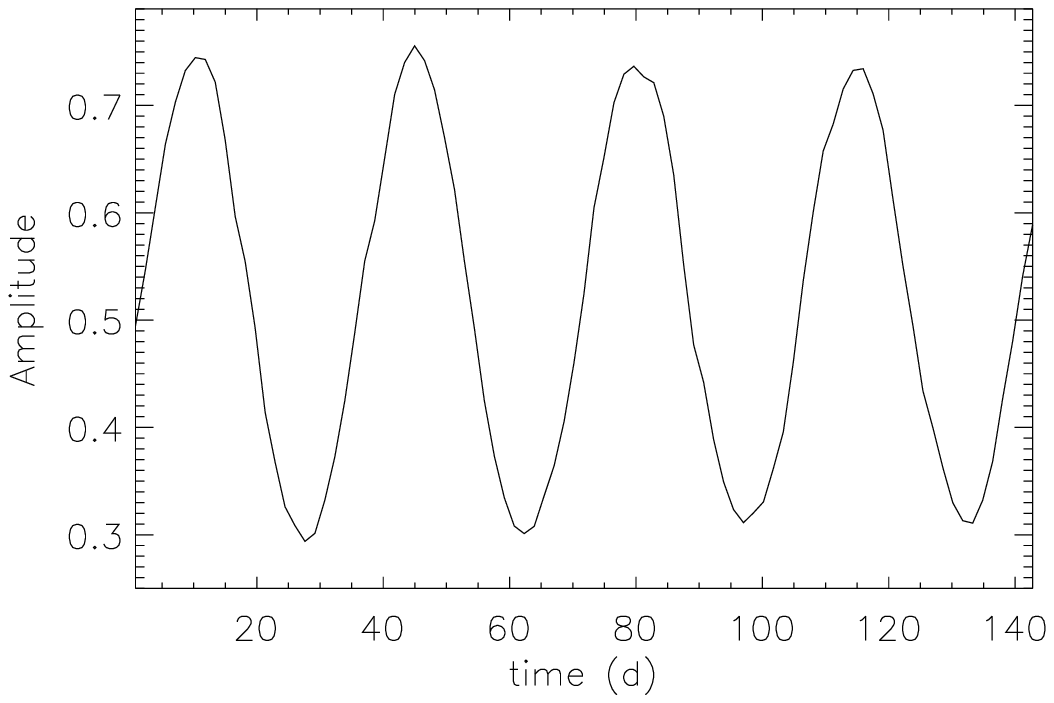}
\caption{
Example of a selected LC (top left panel) and a discarded LC (top right panel),
according to their amplitude variation patterns (bottom panels). Solid red curves are their harmonic fits.
CoRoT ID 105288363 is a Blazhko RR Lyrae star (Guggenberger et al.~\cite{gug11}).
}
\label{figselec}
\end{center}
\end{figure}

We suggest that
the highest recovery fraction, as obtained above automatically, does not reach 100\% because for a number of LCs ($\sim$5\%)
the actual period is not the strongest periodogram peak, even for long-term observations.
This may occur in particular when photometric variability can be modeled as two main sinusoids per cycle.
This is the case, for example,
for some multimode pulsators and also
for many rotating stars that display active regions at opposite faces and produce two main dips per lap.
Therefore, this limitation has instrumental and physical origins.

Accordingly, we developed a simple method for minimizing this problem.
Fig.~\ref{figalias} shows a CoRoT LC that was interpreted here to be such a case.
To identify these cases, the phase diagram was always checked for twice the period $P_1$ of the strongest periodogram peak.
When two dimmings had notably different depths in the phase diagram, the true period $P$ was taken to be $2P_1$,
otherwise it was $P_1$.
In this analysis, the harmonic fit of Eq.~(\ref{eq_best_harm}) was obtained for the phase diagram with a fixed period
(thus without the sum on~$i$) and the final period was also refined to the highest amplitude and lowest dispersion
(as explained above).
This method is not a final solution to avoid aliases, but it certainly helps to reduce this
problem, thus, increasing the recovery fraction closer to 100\%.
Finally, for some LCs with more than one type of variability superposed one another
(e.g.,~rotation +~pulsation), we selected the one that better met the criteria described in
Sect.~\ref{semisinus}. Thus, for some cases, another period, $P_i$ or $2P_i$, was selected instead of $P_1$ or $2P_1$.

\subsubsection{Semi-sinusoidal signature}
\label{semisinus}

Based on CoRoT LCs with known rotational modulations,
as, e.g.,~CoRoT-2 (Silva-Valio \& Lanza~\cite{sil11}),
CoRoT-4 (Lanza et al.~\cite{lan09}),
CoRoT-6 (Lanza et al.~\cite{lan11}), and
CoRoT-7 (Lanza et al.~\cite{lan10}),
the semi-sinusoidal signature was defined here by six main criteria.

\begin{enumerate}[(i)]
\item The variability period is longer than $\sim$0.3 days.
\item The mean amplitude is typically $\lesssim$~0.5~mag.
\item The periodogram shows a relatively narrow spread around the variability peak.
\item The flux maximum and minimum per cycle are often asymmetric with respect to the flux average per cycle.
\item The amplitude varies randomly and smoothly, with a characteristic period of $\sim$~10--30$\times$ of the variability.
\item The short-term flux variation has a smooth semi-sinusoidal shape that can be superposed with a second semi-sine, near in period, varying independently and smoothly in amplitude and phase\footnote{In some cases, a faint third semi-sine contribution may also be found.
Superposed semi-sines may be due to spots but also to some pulsations.
The other criteria must be analyzed carefully to validate this one.}.
\end{enumerate}
Such a detailed description is needed because rotational modulation may indeed present very complex patterns
and these criteria avoid subjectivity in the visual inspection.
The ranges in criteria~(i) and~(ii) are those expected for most rotating stars (e.g.,~Eker~et al.~\cite{eke08}; Hartman et al.~\cite{har10})
and were chosen as a compromise to save time.
To apply criteria~(i) and~(ii) the Lomb--Scargle periodogram was calculated within the Nyquist-frequency range for all LCs
and those with the main periodogram peak with a frequency greater than 0.3~c/d or a mean amplitude greater than~0.5~mag were discarded.

\begin{figure*}
\begin{center}
\includegraphics[width=4.3cm]{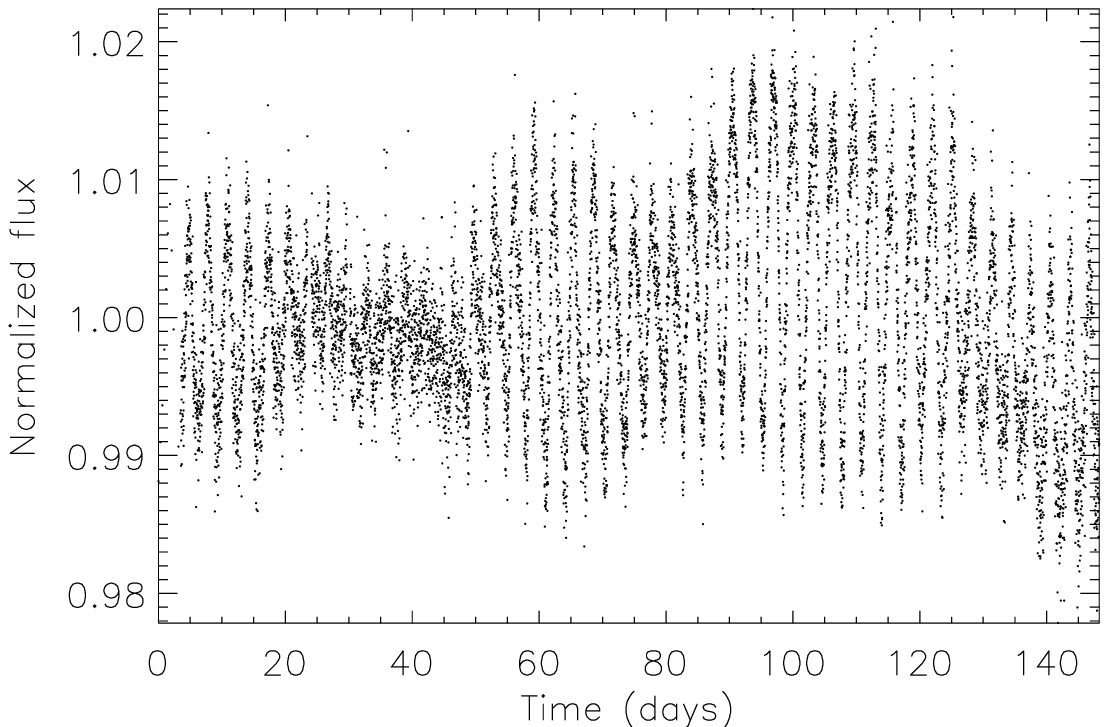}
\includegraphics[width=4.3cm]{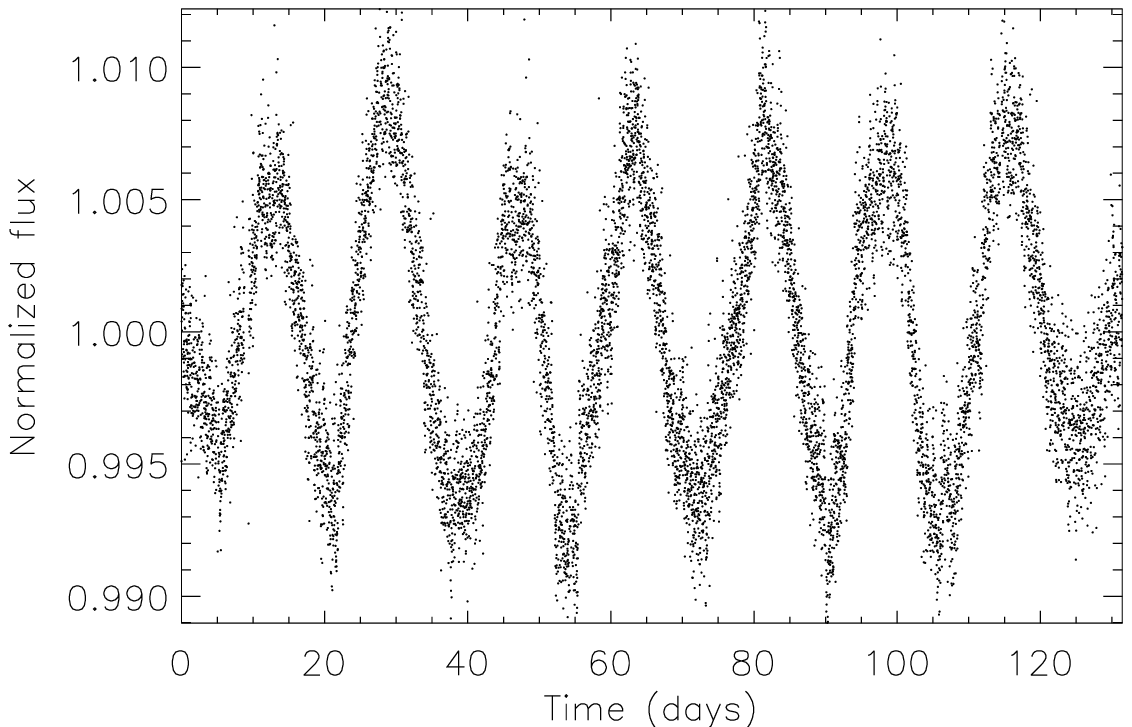}
\includegraphics[width=4.3cm]{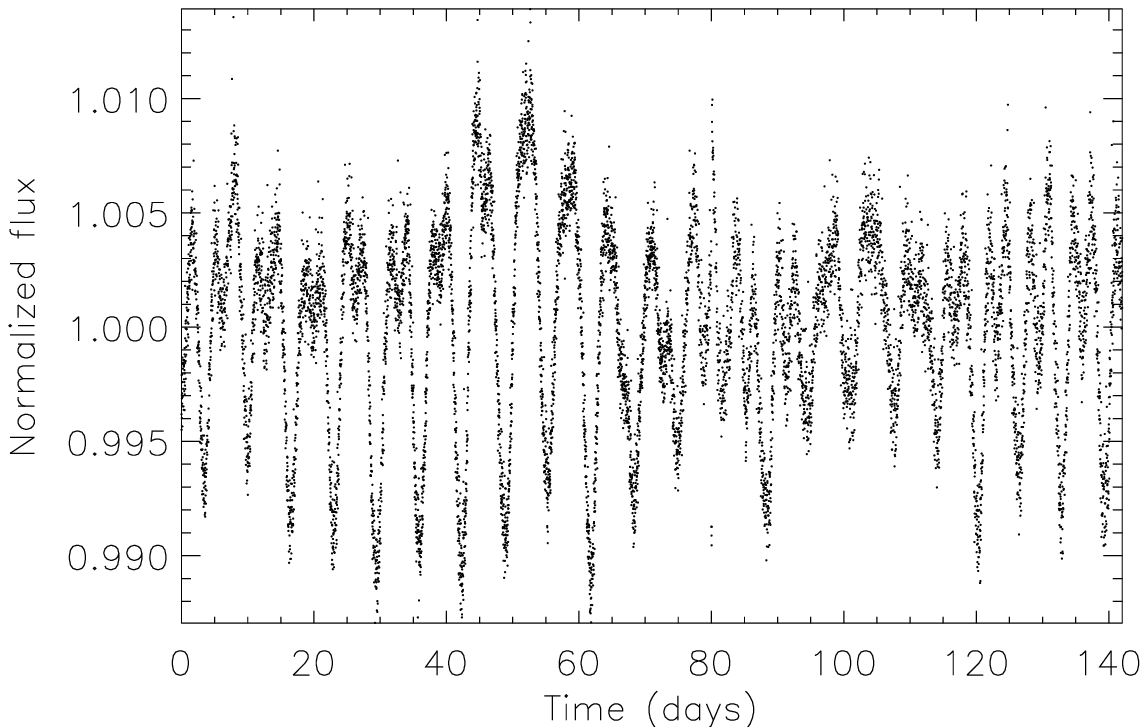}
\includegraphics[width=4.3cm]{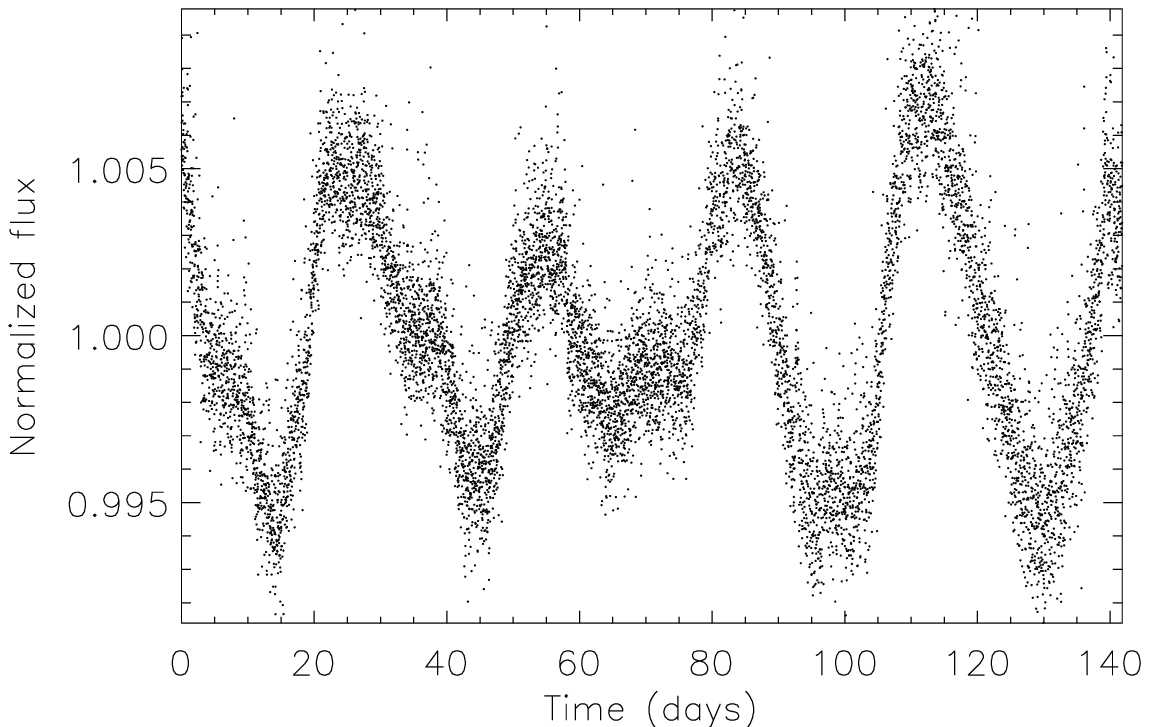}
\includegraphics[width=4.3cm]{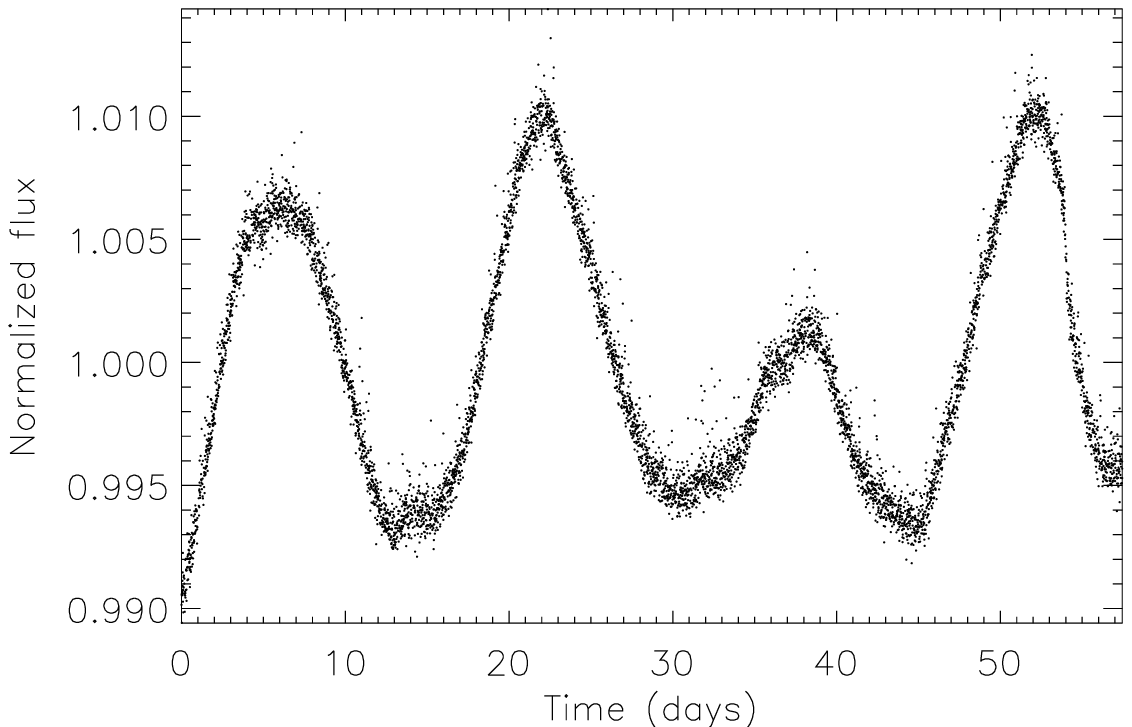}
\includegraphics[width=4.3cm]{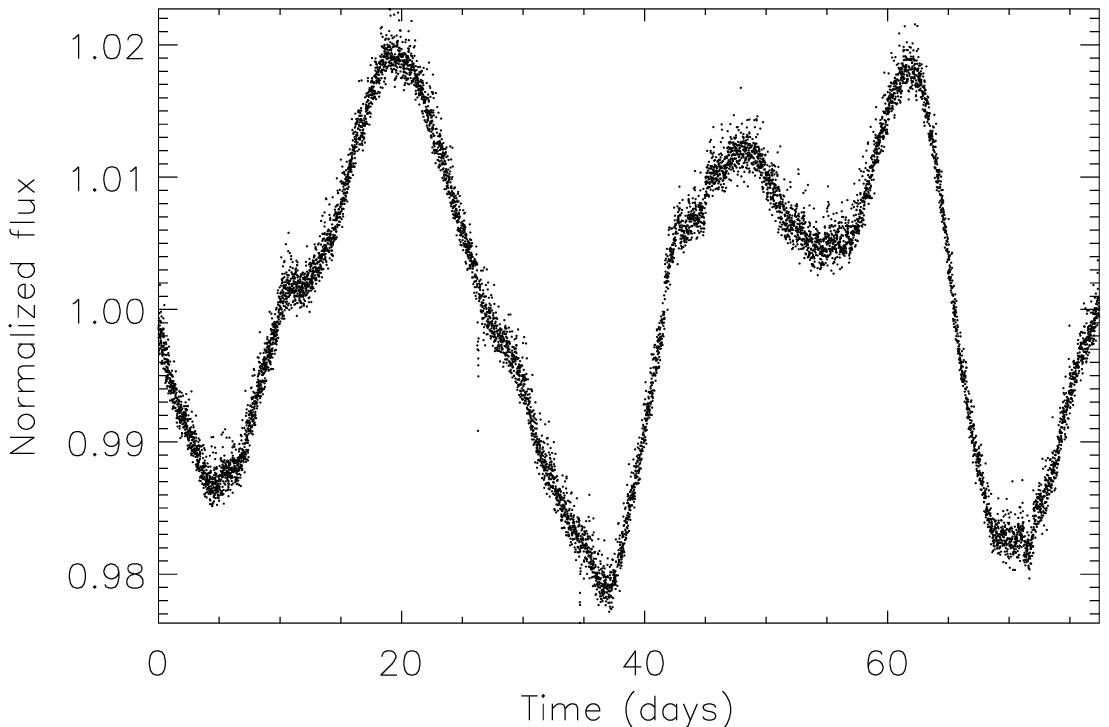}
\includegraphics[width=4.3cm]{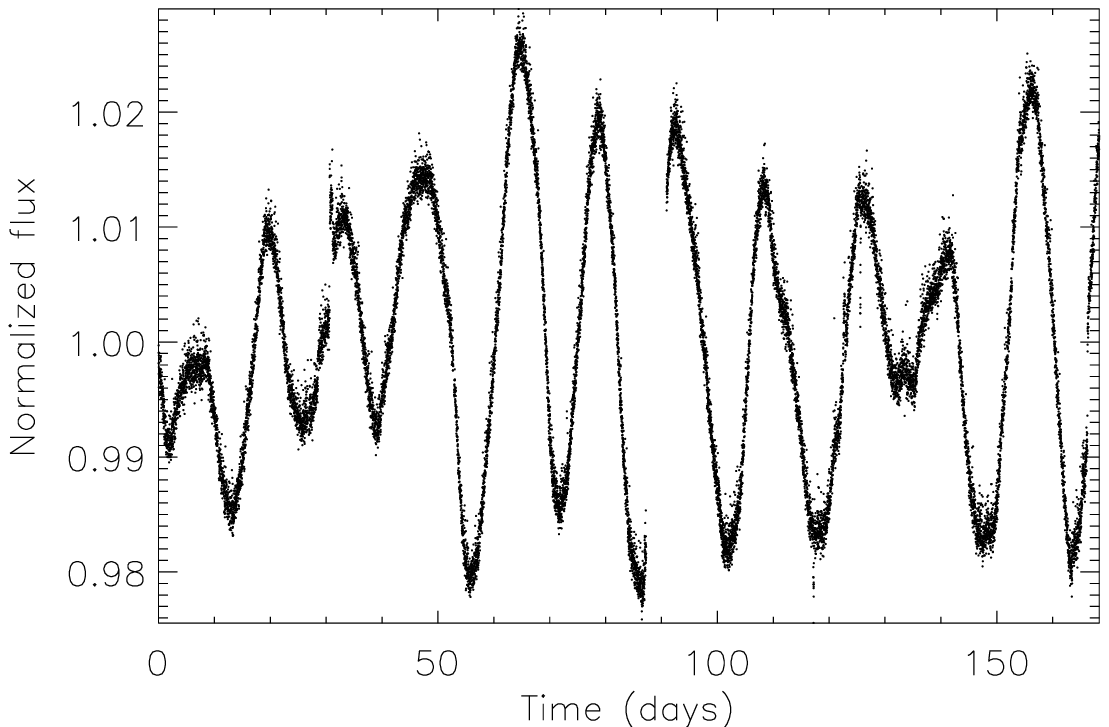}
\includegraphics[width=4.3cm]{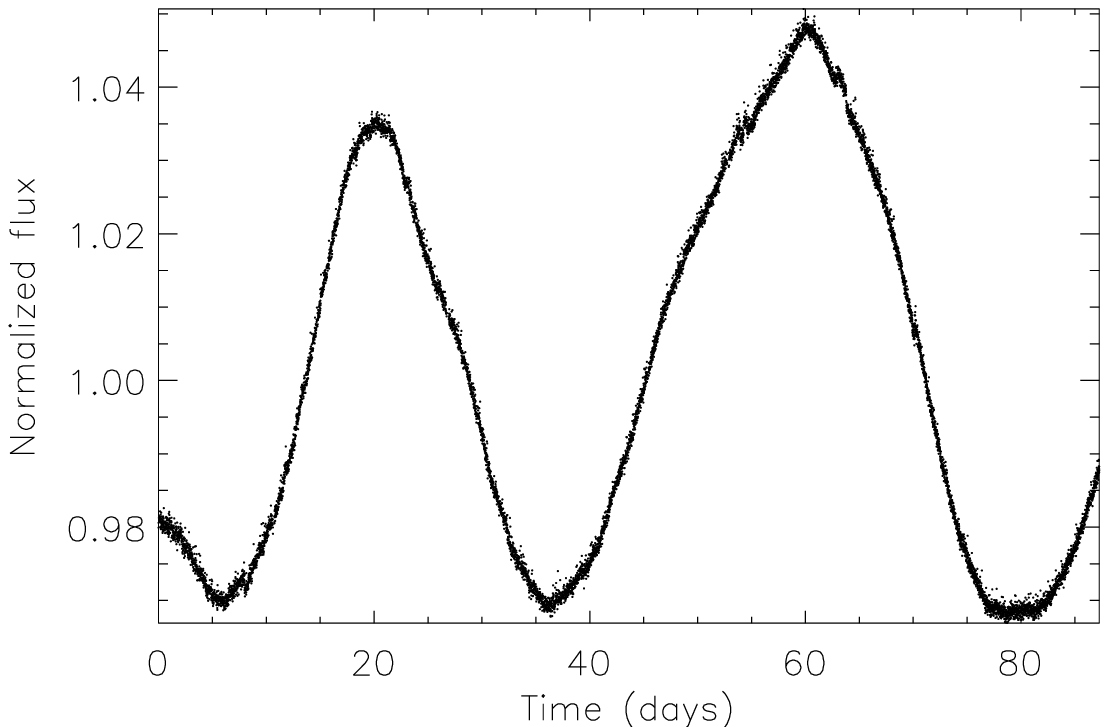}
\includegraphics[width=4.3cm]{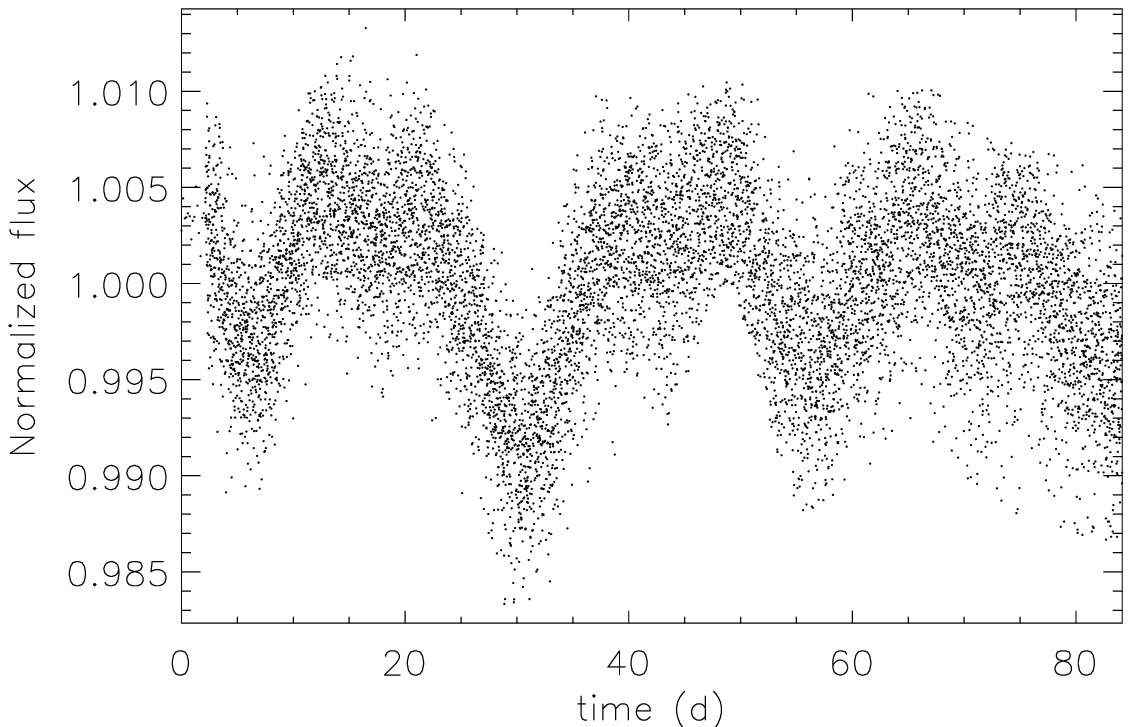}
\includegraphics[width=4.3cm]{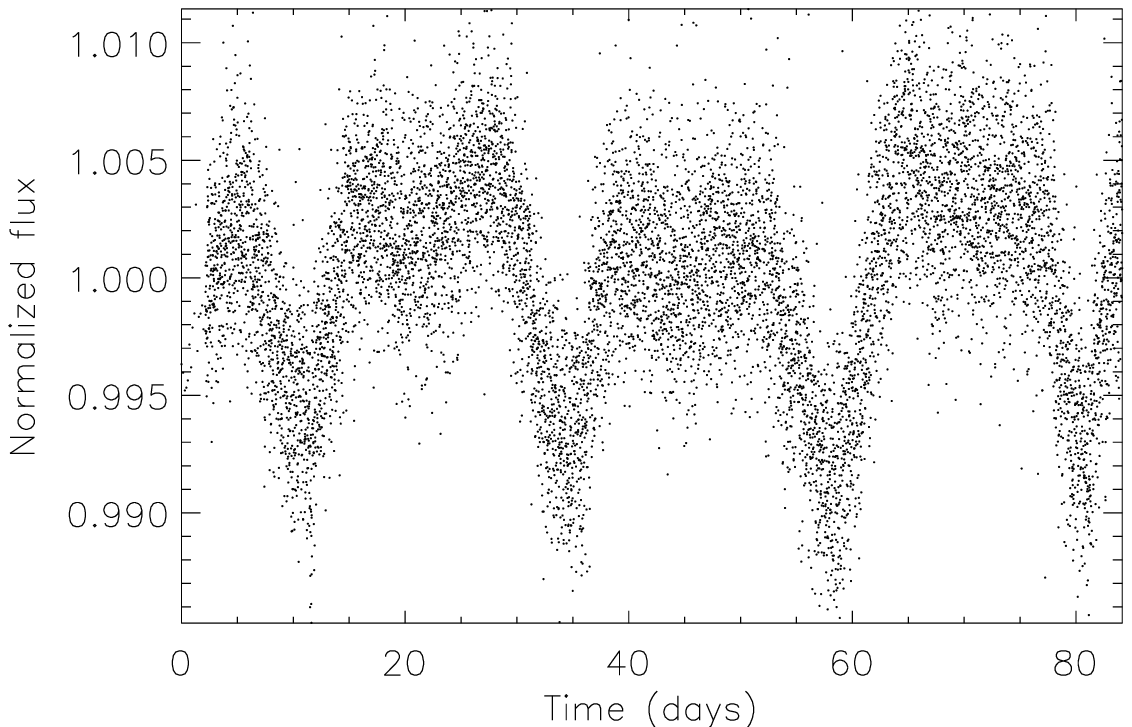}
\includegraphics[width=4.3cm]{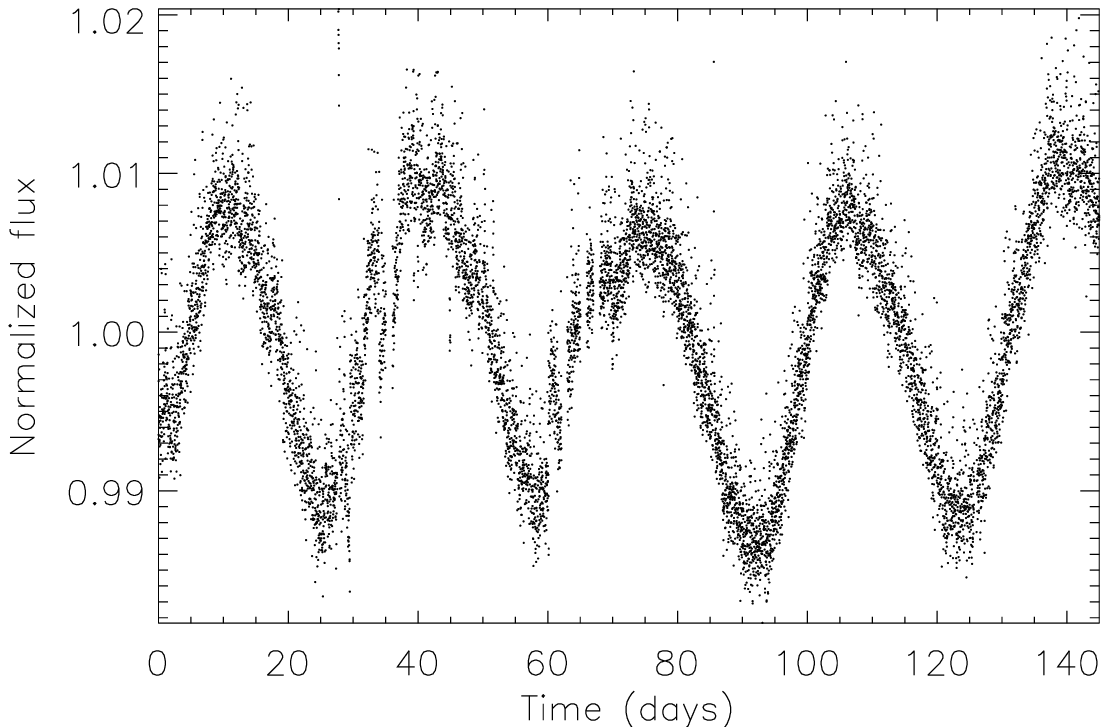}
\caption{
Subsample of LCs presenting the typical variabilities considered in our analysis. The upper panel illustrates LCs of FGK stars, the
middle panel shows LCs of M-type stars, and the lower panel depicts the LCs for the Sun-like candidates defined in Sect.~\ref{subanalogs}.}
\label{figsampling}
\end{center}
\end{figure*}

Criteria~(iii) to~(vi) were applied by visual inspection.
Based on criterion~(iii), variabilities showing strongly spread peaks in the periodogram (e.g.~some semiregular and irregular variables)
were discarded.
Based on criterion~(iv), several pulsators, which usually show nearly symmetric maximum and minimum flux per cycle, were rejected.
Based on criterion~(v),
variabilities with regular amplitude variations (e.g.,~RR Lyrae) or
with nearly constant amplitude (e.g.,~eclipsing binaries) were also rejected.
Fig.~\ref{figselec} shows the example of a selected LC
and its amplitude over time\footnote{These amplitude variations were calculated
within boxes with a duration equal to the variability period.}.
The figure also shows the example of a discarded LC, whose amplitude variations are regular.
Finally, for the LCs we kept, criterion~(vi) was used to select those with a short-term semi-sinusoidal behavior.
Note that the CoRoT time window and noise limits in many cases hampered a proper analysis of criteria~(iii) to~(vi) altogether.
In particular, the long--period variablities were more often subject to a mis-selection
(and these were often classified as the lower confidence group defined in Sect.~\ref{perdet}), but they were still selected here because of their importance in studying stellar evolution.
Therefore, considering that the final sample is a list of candidates, the selection was not very conservative.

Instrumental effects were taken into account using
a procedure similar to that described in Degroote et al.~(\cite{deg09}).
We selected 1000 LCs of different runs, interpreted as constant stars,
and computed the average of their Fourier periodograms to identify instrumental signatures.
The variabilities found in each individual LC were visually compared with the instrumental signatures and periodicities identified as instrumental were rejected.

It is important to note that identifying LCs with semi-sinusoidal signatures as defined here
is useful for selecting rotating candidates if no other information than photometry is available.
However, not all semi-sinusoidal LCs are necessarily produced by rotation and not all rotating variables
produce semi-sinusoidal signatures.
A better selection of rotating variables can only be made with the aid of spectroscopic data.
Nevertheless, selecting this particular type of variation may provide a good filtering of rotating candidates.

\subsubsection{Final selection}
\label{finalselec}

All CoRoT N2 LCs of the exofield were first analyzed automatically to select sources with valid flux measurements, a mean S/N
greater than 1.0 (see Sect.~\ref{snrsel}), and meeting criteria~(i) and~(ii).
Based on 2MASS infrared photometric data, sources from this sample showing contamination and confusion flags were excluded. These flags indicate that photometry and/or 2MASS position measurements of a source may be contaminated or biased by the proximity of an image artifact or a nearby source of equal or greater brightness.

With visual inspection, all methods and criteria described above yielded a final sample of 4,206 targets exhibiting confident semi-sinusoidal variability, as we show in the next section, with spectral types F, G, K, and M and luminosity classes III, IV, and V as listed in the CoRoTSky database.
A portion of the sample has unknown spectral types and luminosity classes, hereafter represented by a question mark~(?).
Table~2, presented in electronic format, displays the computed periods and amplitude of variability, and different stellar parameters (CoRoT ID, right ascention, declination, spectral type, luminosity class, B magnitude, V magnitude, CoRoT run, J magnitude, H magnitude, Ks magnitude, variability period, variability amplitude, and signal-to-noise ratio).
The error average of the variability period is $\sim$3\% and that of the amplitude is $\sim$2~mmag.
Fig.~\ref{figsampling} shows a sample of LCs presenting the typical variabilities considered in our sample, namely a semi-sinusoidal behavior.


%
\subsection{Sample description and biases}
\label{subdescription}

Considering that we had obtained a list of rotating candidates, the
selection methods described above may have biased our sample, for example, by
excluding some regular sinusoidal variabilities.
On the other hand, the selection may
have polluted the final sample with other variables that are not rotators (for instance some semi-regular pulsators)
that may show variabilities somewhat similar to
the semi-sinusoidal signatures.
Of course, because the aim of the methods was to minimize such a sample pollution,
a compromise with some bias is unavoidable.

\begin{figure}
\begin{center}
\includegraphics[width=4.3cm]{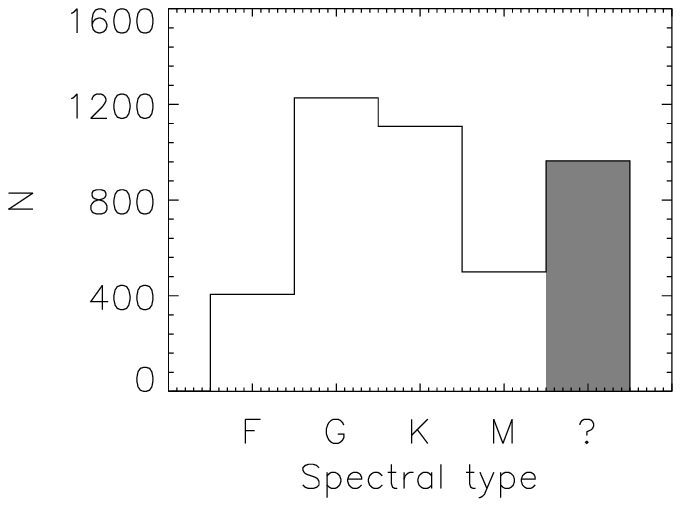}
\includegraphics[width=4.3cm]{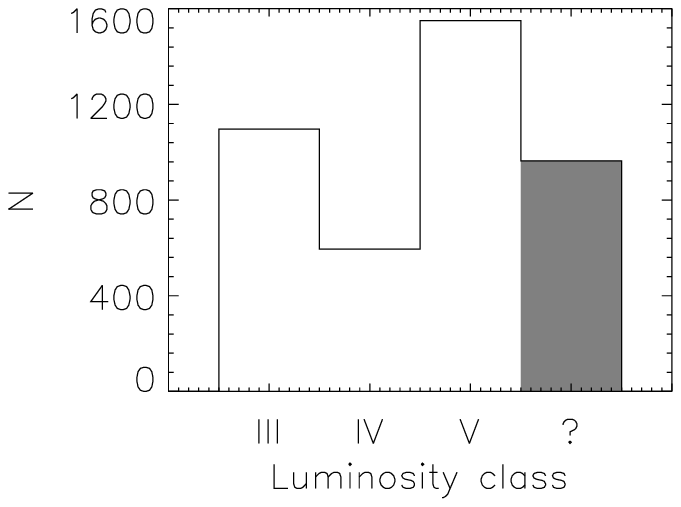}
\caption{
Distributions of spectral types and luminosity classes for the final sample of 4,206 stars analyzed in the present study.} 
\label{fighists}
\end{center}
\end{figure}

\begin{figure}
\begin{center}
\includegraphics[width=4.3cm]{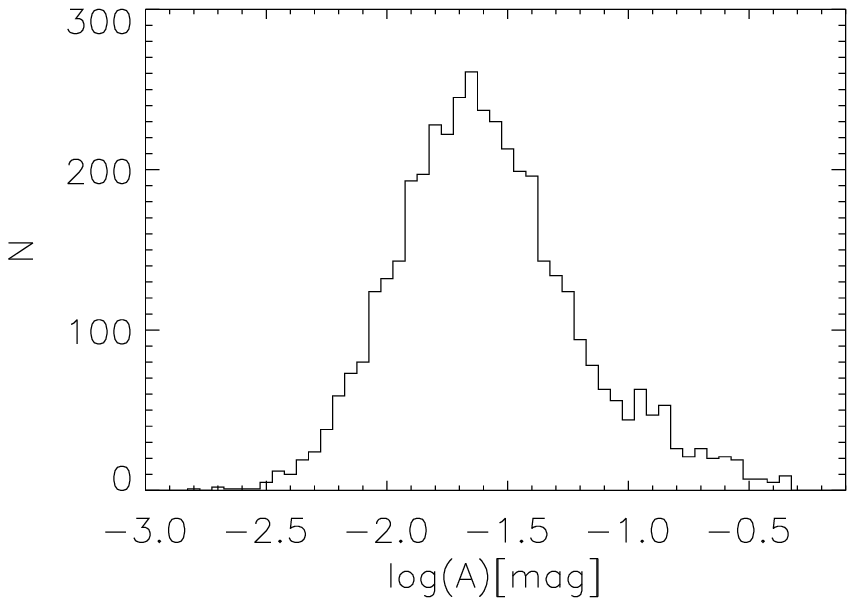}
\includegraphics[width=4.3cm]{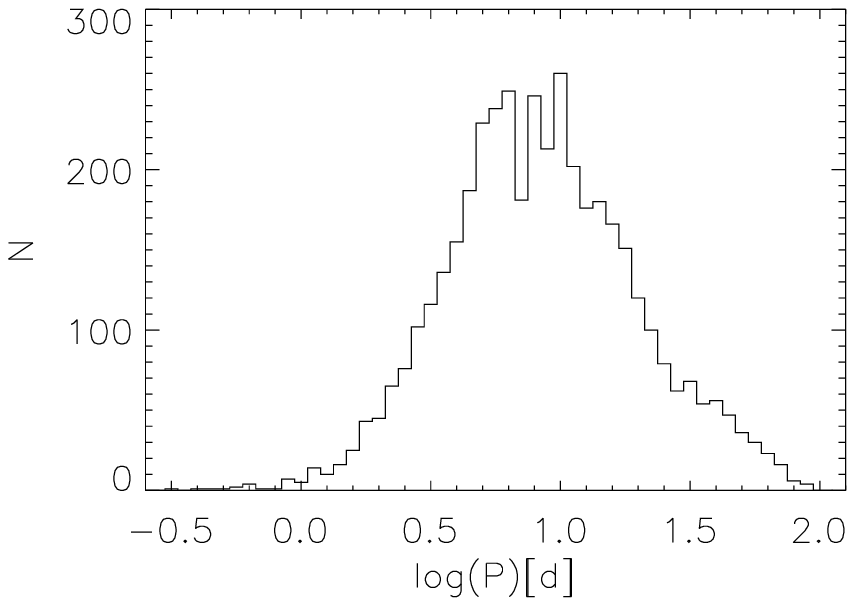}
\caption{
Distribution of variability amplitudes (left panel) and periods (right panel) for the final 4,206 LCs investigated in this study.}
\label{fighists2}
\end{center}
\end{figure}

For a general description of our final sample of 4,206 stars,
Fig.~\ref{fighists} shows their spectral type and luminosity class distributions, while
Fig.~\ref{fighists2} depicts the variability amplitudes in mag and the periods of the corresponding 4,206 LCs.
Thus, most of the stars in our sample exhibit variability amplitudes lower than 0.05 mag, within a range compatible with rotational modulation. However, other types of variabilities may also be found within this amplitude range.
The period distribution (Fig.~\ref{fighists2}, right panel) may include physical aspects,
but they mostly denote biases.
This can be explained by at least two facts.
First, the limited time span of CoRoT LCs of up to $\sim$150~days
makes it more difficult to identify the criteria described in Sec.~\ref{semisinus}
the longer the periods are.
Second, the higher the frequency, the lower the number
of flux measurements cycle by cycle in an LC,
which also complicates identifying
those criteria.
Therefore, in view of the full-width at half-maximum (FWHM) of the period distribution,
the best selection of semi-sinusoidal variabilities in our sample
lies around 3--20~days.

\begin{figure}
\begin{center}
\includegraphics[width=8cm]{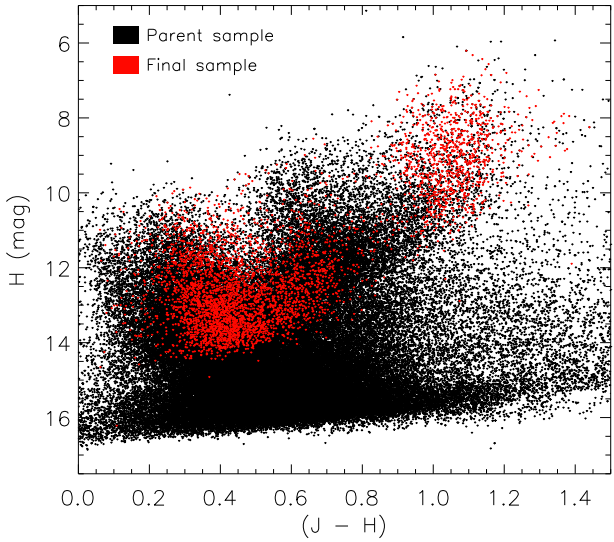}
\caption{
Color-magnitude diagram displaying the (J~--~H) color index versus H magnitude
for the parent CoRoT sample of 124,471 LCs (in black) and for our final sample of 4,206 stars (in red).}
\label{fighr}
\end{center}
\end{figure}

Fig.~\ref{fighr} shows the color-magnitude diagram
by comparing the CoRoT parent sample of 124,471 LCs (in black) with our final sample of 4,206 stars (in red).
This comparison indicates that some more biases were introduced by our selection procedure.
Essentially, there is a cut-off region for stars fainter than~$\sim$14~mag for (J~--~H)~$\lesssim$~0.8
and than~$\sim$11~mag for (J~--~H)~$\gtrsim$~0.8,
caused by the S/N selection described in Sect.~\ref{snrsel}.
We verified that the distribution of the parent sample of 124,471 LCs is quite similar to a
random selection of 2MASS sources.
This means that our sample is valid for relatively bright field stars.
Considering all biases, we are aware that our final sample is not statistically complete. Nevertheless, it is large enough for a robust global analysis if one is cautious to interpret how the biases may affect the physical results.
For example, restricting the sample to that region of the color-magnitude diagram may be advantageous,
because it produces a stronger relation of the color with the stellar evolutionary stage (see Sect.~\ref{results}).

\subsection{Our sample selection versus automatic classifiers}
\label{subdeb09}

It could be suggested that the sample provided in Debossher et al.~(\cite{deb07}, \cite{deb09})
contains all parameters needed for the results presented in Sect.~\ref{results}.
However, their sample was obtained from a fully automatic classifier, which
is useful for the preliminary selection of a large sample of LCs,
but may present a number of problems, particularly for CoRoT LCs, as detailed below.
The importance of our sample
compared with that of Debossher et al.~(\cite{deb07}, \cite{deb09})
for the study of CoRoT targets is justified below.

\begin{figure}
\begin{center}
\includegraphics[width=7cm]{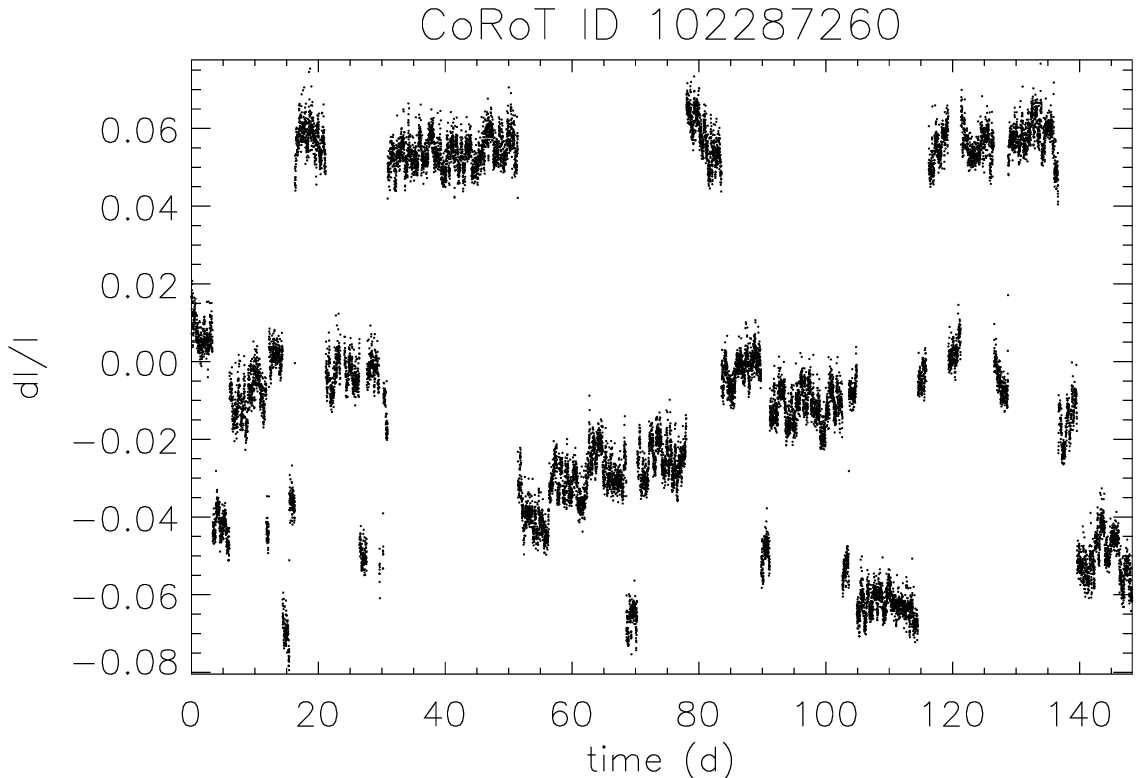}
\includegraphics[width=7cm]{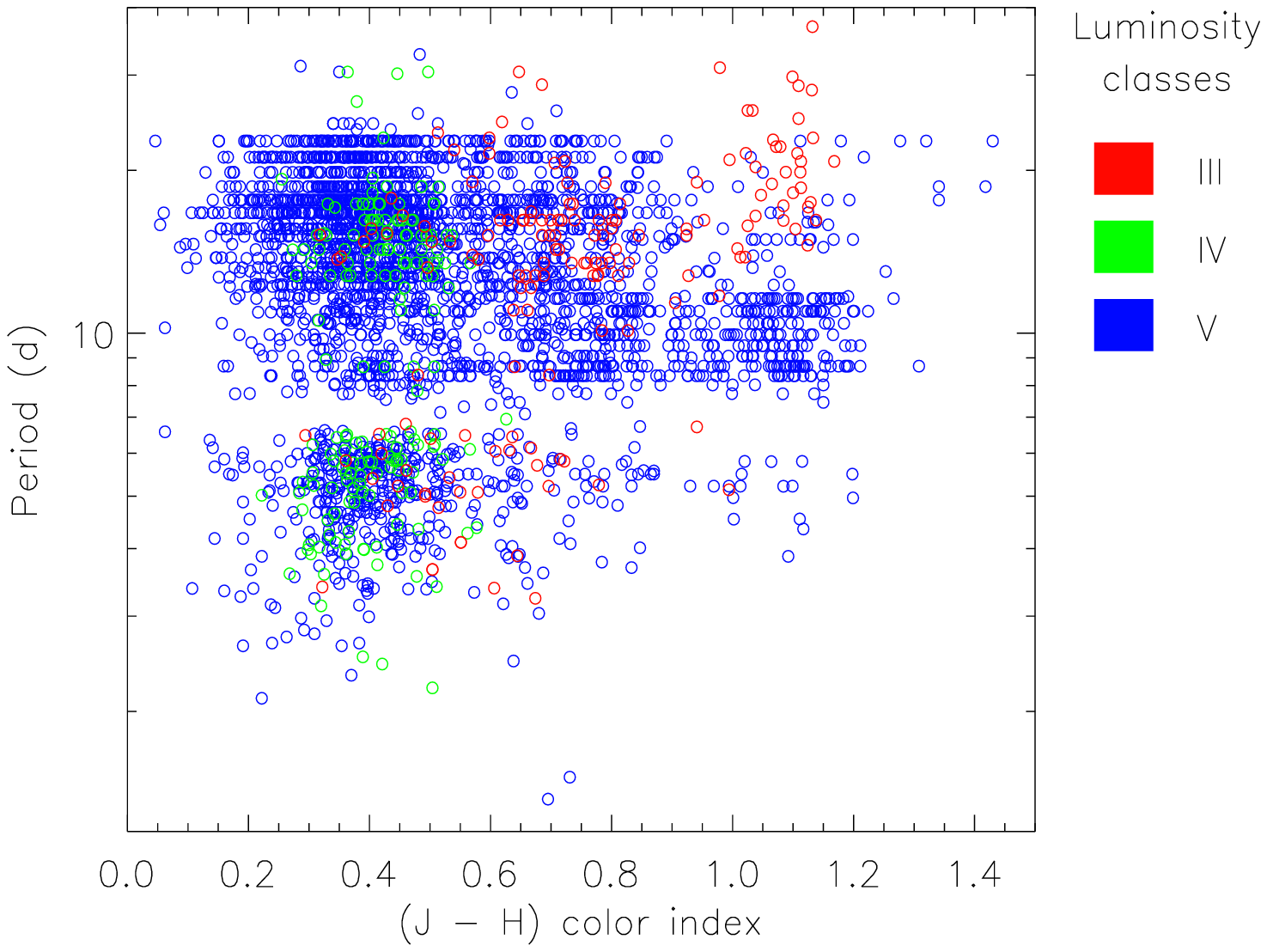}
\caption{
Top: example of a LC automatically misclassified in Debosscher et al.~(\cite{deb07}, \cite{deb09}) as exhibiting rotational modulation with a time period of 6.15 days.
Bottom: color-period distribution for a subsample of stars automatically classified in Debosscher et al.~(\cite{deb07}, \cite{deb09}) and Sarro et al.~(\cite{sar09}) as displaying possible rotational modulation. 
}
\label{figmiscl}
\end{center}
\end{figure}

As mentioned in the introduction, automatic classifiers are subject to misclassifications as a result of data artifacts.
Discontinuities found in CoRoT LCs may be interpreted as variabilities
by producing an incorrect calculation of periods and the statistical measurements used in the classifiers.
For example, Fig.~\ref{figmiscl} (top) shows a CoRoT LC classified by Debossher et al.~(\cite{deb07}, \cite{deb09}) and Sarro et al.~(\cite{sar09}) as displaying rotational modulation with a period of 6.15 days, Mahalanobis distance of 1.36, and class probability of 98.8\%. Although these values are typical of a good classification, visual inspection clearly shows this is a fake period caused by strong discontinuities.

Incorrect classifications as that in Fig.~\ref{figmiscl} (top) may contaminate a sample of stars and therefore hamper the identification of physical results.
For instance,
consider a subsample of FGKM stars classified in Debossher et al.~(\cite{deb07}, \cite{deb09}) and Sarro et al.~(\cite{sar09}) as having rotational modulation, to be compared with our results (see Sect.~\ref{subvarbeh}).
Only the best classifications with Mahalanobis distances smaller than 1.5 and class probabilities higher
than 90\% were considered in that subsample, but it
shows no clear behavior in a color-period diagram, as seen in the bottom panel of Fig.~\ref{figmiscl}.
This is not the case of our sample, which is a source of relevant physical results to be demonstrated in this study.
Visual inspection was crucial in minimizing misclassifications in our sample.

Finally, all CoRot targets classified in Debosscher et al.~(\cite{deb07}, \cite{deb09}) as possible rotating variables were visually inspected by us, independently of their Mahalanobis distances and class probabilities.
Only $\sim$4\% of those targets show semi-sinusoidal signatures
as defined in Sect.~\ref{semisinus}.
Therefore only this fraction was included in our sample.
These targets correspond to $\sim$60\% of our whole sample,
the remaining targets was not classified in Debosscher et al.~(\cite{deb07}, \cite{deb09}) as possible rotating variables.
Therefore, our sample has a substantial number of additional candidates for the study of stellar rotation.
We emphasize that the 96\% of the LCs classified in Debosscher et al.~(\cite{deb07}, \cite{deb09}) that are
not included in our sample are not necessarily misclassifications. The large fraction of rejected LCs, in the context of the present study, indicates that 
these LCs do not fulfill the main criteria adopted by our selection procedure for a semi-sinusoidal signature.

\subsection{Comparison with period measurements available in literature}
\label{subpercomp}

\begin{figure}
\begin{center}
\includegraphics[width=8cm]{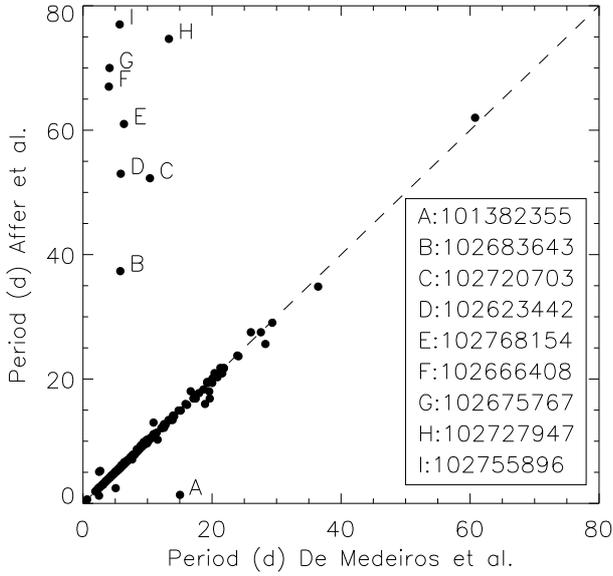}
\caption{
Comparison between period measurements of the present work and those obtained by Affer et al.~(\cite{aff12}).
The most strongly discrepant cases are marked by letters and their CoRoT IDs are listed in the legend.}
\label{figpercomp}
\end{center}
\end{figure}

The literature offers now a substantial list of 1978 period measurements computed from CoRoT LCs, 1727 of which interpreted by their authors as rotation periods (Affer et al.~\cite{aff12}). From this sample, 216 targets are in common with our sample, which offers the possibility for a 
preliminary comparison between the two sets of measurements. Fig.~\ref{figpercomp} displays our period estimates versus those obtained by Affer et al.~(\cite{aff12}). Of the 216 targets in common, the periods agree excellently for about 95\%. For the main discrepant cases, indicated in Fig.~\ref{figpercomp}, the following aspects should be outlined: case~A has two types of variabilities superposed, one of which was selected by us and the other by Affer et al.~(\cite{aff12}). The variability selected by us is compatible with a semi-sinusoidal signature,
while the latter has an amplitude approximately constant over time, which is more often observed in eclipsing binary LCs (see Sect.~\ref{semisinus}).
For the other cases (B to I), our periods match the semi-sinusoidal signature, while the periods given by those authors correspond to long-term contributions not compatible with such a signature. A relevant aspect of this comparative analysis is that, except for case~A, the disagreement is associated with the long period measurements computed by Affer et al.~(\cite{aff12}).


\subsection{Influence of reddening}
\label{subreddening}

We used the (J~--~H) color index obtained from the 2MASS photometry in our analysis,
which may be affected by reddening.
To determine its effect on our results, we computed pseudo-colors as described in Catelan et al.~(\cite{cat11}), which are supposed to be reddening-free.
For instance, these authors considered the data collected with a set of five different broadband filters
of the Vista Variables in the V\'ia L\'actea (VVV) ESO Public Survey
to estimate reddening-free indices, which can be calculated by calibrating magnitudes or colors. 
Accordingly, we used 2MASS magnitudes and equation (7) in Catelan et al.~(\cite{cat11})
to determine the pseudo-colors. As demonstrated below, a reddening
correction can dramatically affect the behavior in the period-versus-color distribution for our
stellar sample.

\section{Results and discussion}
\label{results}

The aim of this pioneering investigation is to identify and quantify the level of 
semi--sinusoidal variability in stellar LCs produced by the CoRoT space mission.
To that end, we dedicated most of our effort to identifying through visual inspection the LCs without ambiguities in their semi--sinusoidal behavior. As a result, 4,206 periods of variability for stars of spectral types F, G, K and M are now available. This section presents some statistics and 
characteristics of the periods obtained, in particular as a function of colors.

\subsection{General description of the variability behaviors}
\label{subvarbeh}

\begin{figure}
\begin{center}
\includegraphics[width=\columnwidth,height=7cm]{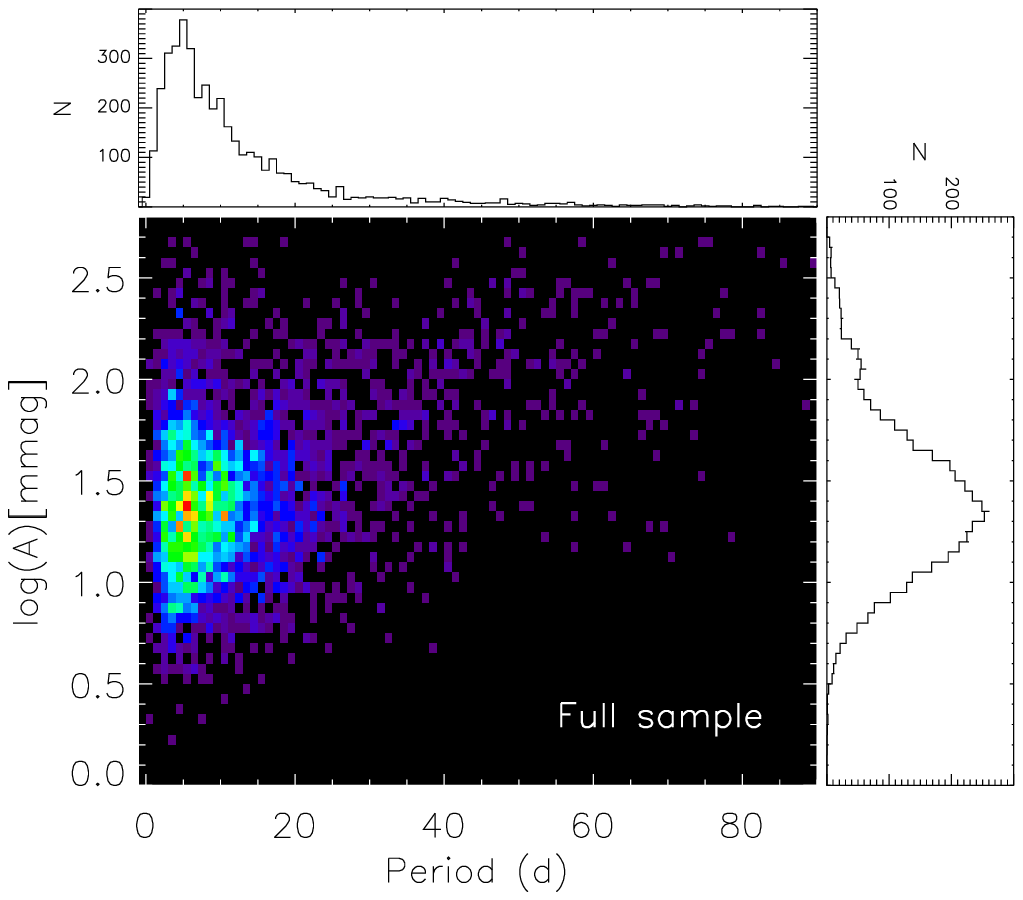}
\includegraphics[width=\columnwidth,height=7cm]{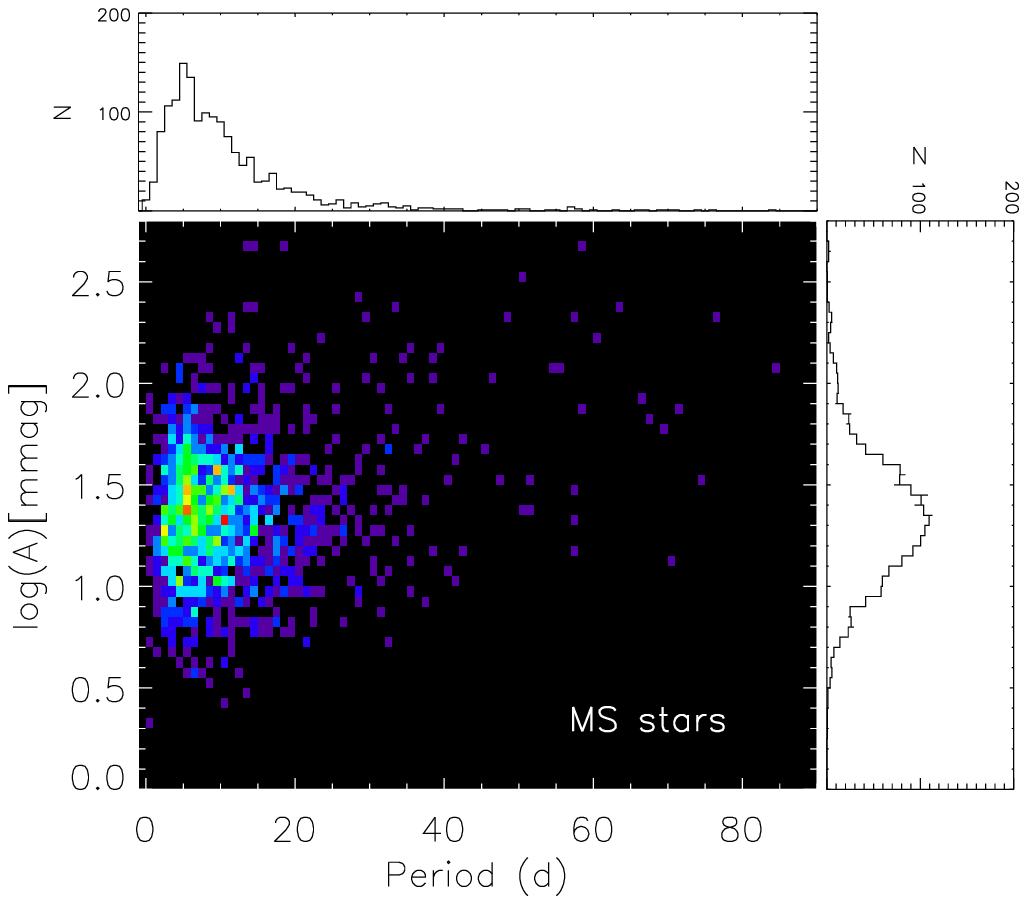}
\caption{
Logarithm of variability amplitude (in mmag) as a function of period.
Top panel: the final sample described in Sect.~\ref{secobs}. Bottom panel: only stars from luminosity class V.
In the colored grids, red represents a greater number of stars, purple indicates a lower number of stars, and black depicts an absence of stars.
The distribution of amplitude is shown at the right side of each panel and the period distribution is indicated at the top of each panel.
}
\label{figampliper}
\end{center}
\end{figure}

Fig.~\ref{figampliper} (top panel) shows the variability amplitude as a function of variability period for the final sample of 4,206 selected stars.
One also observes a slight trend of finding higher amplitudes at longer periods.
Whether this has a physical contribution or is only caused by biases is not clear.
In principle, the longer the period, the more difficult it is to detect a faint signal --
because of the fewer observed cycles --
which could produce a bias in the amplitude.
However, the cut-off by S/N based on Fig.~\ref{figsnr}
should have reduced this possible bias.
On the other hand, the observed behavior for the period dependence on amplitude in
Fig.~\ref{figampliper} may also be affected by a color bias (see Fig.~\ref{fighr}).
Despite these biases,
this behavior may still have a physical influence, as we discuss in the beginning of the next section.

Fig.~\ref{figampliper} (bottom panel) depicts variability amplitude as a function of the variability period for a subsample of main-sequence stars (selected from CoRoT luminosity class).
According to Basri et al.~(\cite{bas11}), main-sequence stars are expected to be more active for shorter periods and may be more obviously periodic or display larger variation amplitudes (unless activity was too uniformly distributed).
Therefore, one could expect some trend for an amplitude decrease with increasing period
for main-sequence rotating variables, which is not observed in our sample.
However, according to literature data, this behavior was also absent from a sample of field stars selected as rotating candidates in Hartman et al.~(\cite{har10}) and in a sample of chromospherically active binaries with photometric rotation periods studied in Eker et al.~(\cite{eke08}).\footnote{More details about those public data are provided in Sect.~\ref{subdiscvar}.}

\begin{figure}
\begin{center}
\includegraphics[width=\columnwidth]{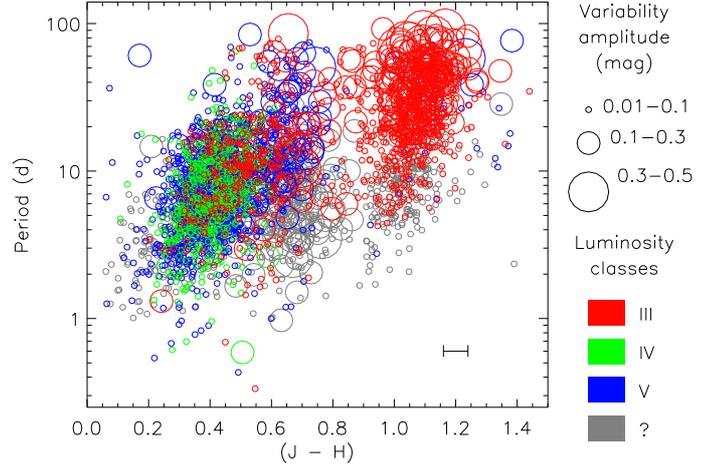}
\caption{
Color-period diagram without reddening correction, demonstrating the variability period as a function of the color index (J~-~H) for the final sample described in Sect.~\ref{secobs}.
Circle size indicates the variability amplitude in mag and colors represent the luminosity class.
The typical error of (J~--~H) is displayed in the error bar.
}
\label{figpercolor}
\end{center}
\end{figure}

\begin{figure}
\begin{center}
\includegraphics[width=\columnwidth]{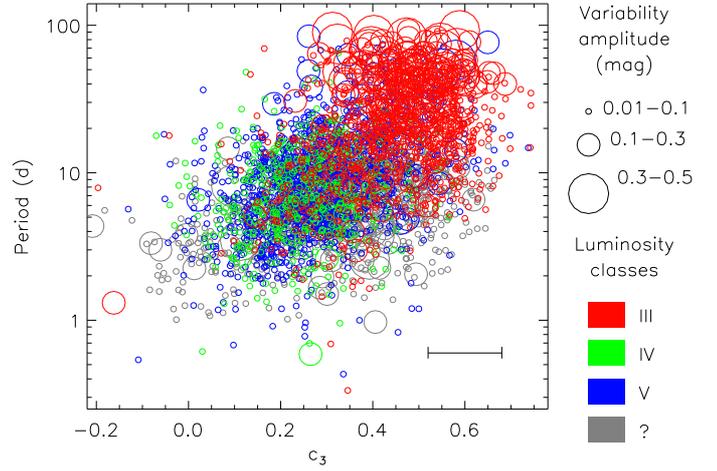}
\caption{
Color-period diagram, showing the variability period as a function of a pseudo-color for our final sample described in Sect.~\ref{secobs}. This pseudo-color, denominated $c3$, is suggested as reddening-free by Catelan et al.~(\cite{cat11})
and is computed as described in Sect.~\ref{subreddening}.
Circle size represents the variability amplitude in mag and colors indicate the luminosity class.
The typical error of $c_3$ is displayed by the error bar.
}
\label{figpercolwrc}
\end{center}
\end{figure}

By combining period, color index, amplitude, and luminosity class, we achieved a detailed overview of the
semi-sinusoidal variabilities for the sample described in Sect.~\ref{secobs}.
Fig.~\ref{figpercolor} shows the color-period diagram, where the variability period is plotted as a function of color index (J~--~H).
Circle size indicates the variability amplitude in mag and colors correspond to the CoRoT luminosity class.
This figure provides outline variabilities in the evolutionary context.
From a global perspective, 
there are at least two important facts in this color-period diagram.
First, it shows two distinct stellar populations: one to the left, with (J~--~H) $\lesssim$ 0.85, and another to the right, with (J~--~H) $\gtrsim$ 0.85.
Second, these populations tend to show an increase of the period with increasing color index, each at a different rate.
These two distinct populations should be related to different evolutionary stages of the stars.
In fact, there is a substantial number of giant stars in the population to the right.
On the other hand, stars from classes III, IV, and V are more or less uniformly distributed to the left.
This dispersion of luminosity classes may be associated with uncertainties in the parameters of the CoRoTSky database.
There are also quite a few stars with low amplitude variability in the region with (J~--~H) $<$ 0.55,
while those with a color index between 0.55 to 0.9 mostly have higher amplitudes.

Fig.~\ref{figpercolwrc} shows the color-period diagram, where, instead of (J~--~H), we used
a pseudo-color, $c_3 = (J-H) - 1.47 (H-K_S)$,
suggested to be reddening-free in Catelan et al.~(\cite{cat11}, Sect.~2.4). In this case, the gap that was
clearly seen in Fig.~\ref{figpercolor} is not evident, even though close
inspection of the distribution confirms that the $c_3$ distribution
is also bimodal, with the two modes strongly tied to the same
two modes that are seen in Fig.~\ref{figpercolor}. There are at least three possible
reasons for the less prominent gap in Fig.~\ref{figpercolwrc}, as compared to Fig.~\ref{figpercolor}.
First, the combination of three filters in the case of $c_3$, as opposed
to just two in the case of (J~--~H), leads to an increase in the propagated
errors, and accordingly, the typical error bars in the $c_3$ values are larger
than the corresponding ones in (J~--~H). Second, theoretical evolutionary
tracks reveal that the extension of the Hertzsprung gap is reduced (in
mag units) when going from (J~--~H) to $c_3$: for instance, for a
$3 \, M_{\odot}$ star, based on evolutionary tracks from the BaSTI
database (Pietrinferni et al.~\cite{pie04}), the interval between the main-sequence
turnoff point and the base of the RGB amounts to about 0.5~mag
in $J-H$, but only about 0.35~mag in $c_3$. Third, (J~--~H) is found to be
more tightly correlated with the spectroscopic temperatures from Gazzano
et al.~(\cite{gaz10}) than $c_3$~-- an effect that may also be related to the
increased errors that affect the latter quantity.

\subsection{Root-cause of the semi-sinusoidal variability in CoRoT LCs}
\label{subdiscvar}

The observed color-period scenario (Figs.~\ref{figpercolor} and~\ref{figpercolwrc}) may be reflecting some physical contribution
even if there is some bias, following the discussion presented
in Sect.~\ref{subdescription}. As observed in Fig.~\ref{fighr}, evolved sources are rather selected at higher colors.
Thus, such a bias may produce an evolutionary selection that can partially explain the behavior
observed in Figs.~\ref{figpercolor} and~\ref{figpercolwrc}.
For example, if our sample is composed of rotating candidates,
it is natural to expect longer periods for higher colors,
based on physical reasons related to stellar evolution theories (e.g.,~Ekstr\"om et al.~\cite{eks12}).
This may also be the case in Fig.~\ref{figampliper} (top panel) for the amplitude.

To check our results,
we compared our sample with $\sim$~1800 field stars available in the HATNet Pleiades Rotation Period Catalogue described in Hartman et al.~(\cite{har10}).
The authors conducted a survey to determine stellar rotation periods in the Pleiades cluster and obtained photometric periods of non-cluster members. The non-cluster members, assumed to be field stars, show variabilities suspected to be rotational modulation, several of which may have other physical natures, as observed in the present study.
For these field stars, the period distribution is very similar to those of our sample,
but there are several ($\sim$35\%) sources containing periods between about 0.1 and 1.0 days.
In the color-period diagram, field stars observed in Hartman et al.~(\cite{har10})
are reasonably compatible with our sample for the $\gtrsim$1.0 day period.
As in our sample, there is a slight increase in amplitude with the rise in color index.

Furthermore,
we analyzed the data available in the catalog of chromospherically active binaries provided in Eker et al.~(\cite{eke08}), which contains information on brightness, colors, photometric and spectroscopic data, and physical quantities for 409 field and cluster binary stars. These data provide a basis for determining to what extent our sample exhibits photometric characteristics similar to those of stars with measured rotation periods. Indeed, some binary systems may be impacted by tidal effects; however, the overall statistics of the sample in Eker et al.~(\cite{eke08}) can be considered for comparison with our sample. Moreover, our sample may also be composed of non-eclipsing binary systems affected by tidal interactions.
The color-period diagram of the Eker et al.~(\cite{eke08}) sample shows higher amplitudes for (J~-~H)~$\gtrsim$~0.55, in line with our sample. Nevertheless, the amplitude range of the sample of Eker et al.~(\cite{eke08}) has a maximum around 0.05~mag, while the highest amplitudes for our sample occur at around 0.025~mag, possibly because CoRoT was designed to observe fainter sources.
In summary, the global behavior of the variabilities in our final sample is compatible in many aspects with that expected for rotating stars, based on the literature. Nevertheless, we should be cautious with the interpretation of the whole list of periods,
because physical phenomena other than rotation can produce LCs with semi-sinusoidal behavior, as discussed here.

\subsection{Rotating Sun--like candidates?}
\label{subanalogs}

The rotation period of the Sun ranges from 23 days at the equator to 33.5 days at the poles (e.g.,~Lanza et al.~\cite{lan03}). Based on solar values of the (J~--~H)$_\odot$ color index defined in the literature, we made an additional effort to identify stars in the present sample with (J~--~H) colors near the solar value that display variability periods close to the Sun's rotation, namely rotating Sun-like stars. According to recent research, the solar (J~--~H)$_\odot$ index ranges from 0.258 (Holmberg et al.~\cite{hol06}) to 0.355 (Rieke et al.~\cite{rie08}). In addition, Zhengshi et al.~(\cite{zhe10}) computed (J~--~H)$_\odot$ = 0.288, from Valcarce et al. (\cite{val12}) we estimated (J~--~H)$_\odot$ = 0.347, whereas Casagrande et al.~(\cite{cas06}) give a list of different estimates of solar color indexes (J~--~H)$_\odot$. Based on these references, we established an average (J~--~H)$_\odot$ = 0.315 $\pm$ 0.04.
Within the solar rotation period from 23 to 34 days and for color indices (J~--~H) between 0.275 and 0.355, we identified two stars; however, only one source is a G-type star of luminosity class V, with an amplitude lower than 0.05 mag.
Considering a (J~--~H) range two times wider, from 0.235 to 0.395, results in a total of three stars exhibiting period and amplitude variability, as well as a spectral type and luminosity class close to the Sun. 
Therefore, one of the by-products of this study is a set of three rotating Sun--like candidates in the context of photometric period, namely the CoRoT IDs 104049149, 104685082, and 105290723.
According to Figs.~\ref{fighists} and~\ref{fighists2}, the identified candidates seem to be compatible with the number of targets available in our sample, considering the distribution of spectral type, color, and variability period.

\subsection{Variability of M-type stars}

Rotational modulation in M-type stars can be considered possible based on the results reported by
H\"unsch et al.~(\cite{hun01}). The authors examined M-type giant stars and found indications of variability in H-alpha and Ca I 6572, which may be related to chromospheric activity.
Our sample contained 96 stars of spectral type MV with amplitude variability ranging between 0.004 and 0.2 mag and 416 stars of spectral type MIII with amplitudes ranging between 0.01 and 0.5 mag.
Of course, follow-up is needed to check the nature of these variabilities,
but this may be a substantial amount of M-type stars with rotational modulation.

Our sample of M-type stars may also be useful for future studies,
based for example on the investigation conducted by Herwig et al.~(\cite{her03}).
These authors analyzed the s-process in rotating stars of the asymptotic giant branch (AGB),
but data obtained to date are insufficient to understand many aspects of stellar evolution.
This is because M-type giant stars -- either RGB or AGB -- generally do not exhibit significant stellar activity.
New results may be obtained if at least a fraction of our M-type stars are confirmed to present rotational modulation.

\subsection{Is there a center versus anti-center difference in the behavior of variability period distributions?}

\begin{figure}
\begin{center}
\includegraphics[width=\columnwidth,height=3cm]{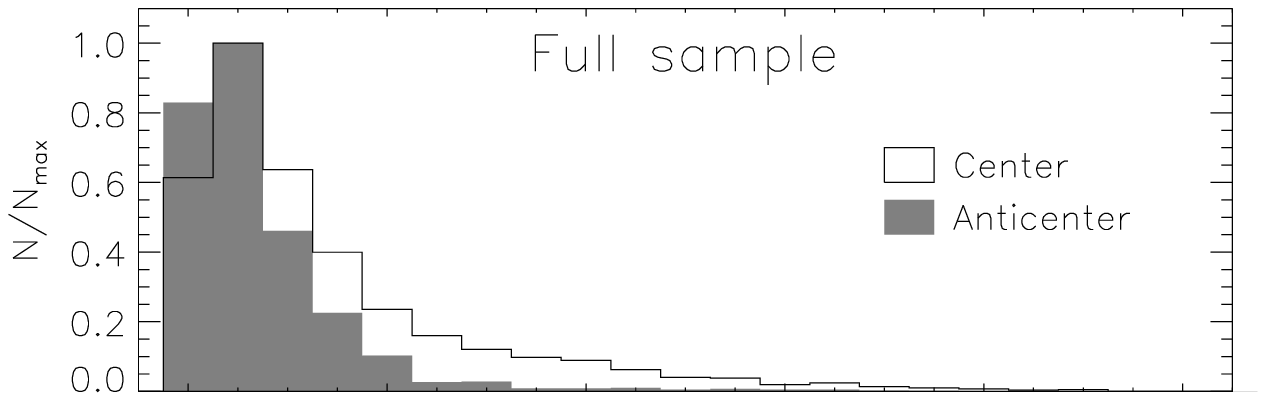}
\includegraphics[width=\columnwidth,height=3cm]{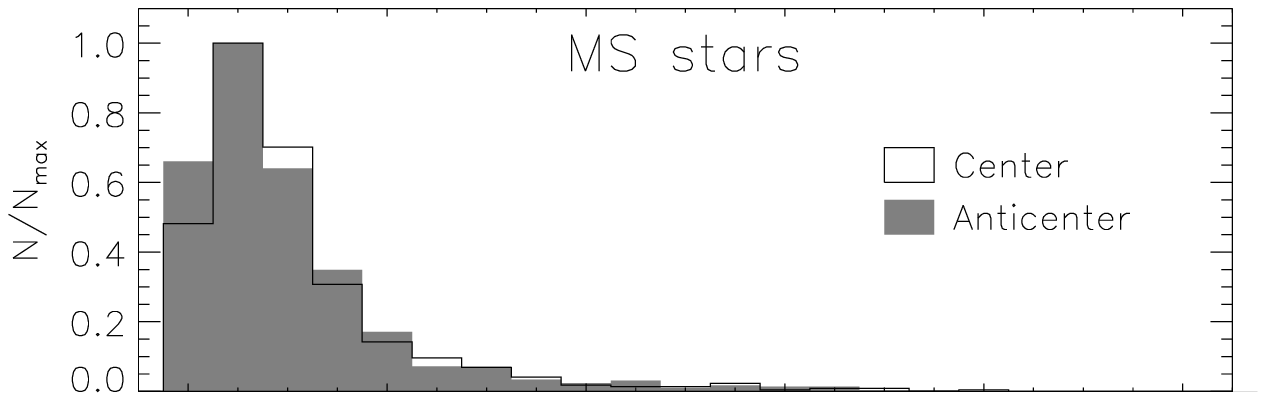}
\includegraphics[width=\columnwidth,height=3cm]{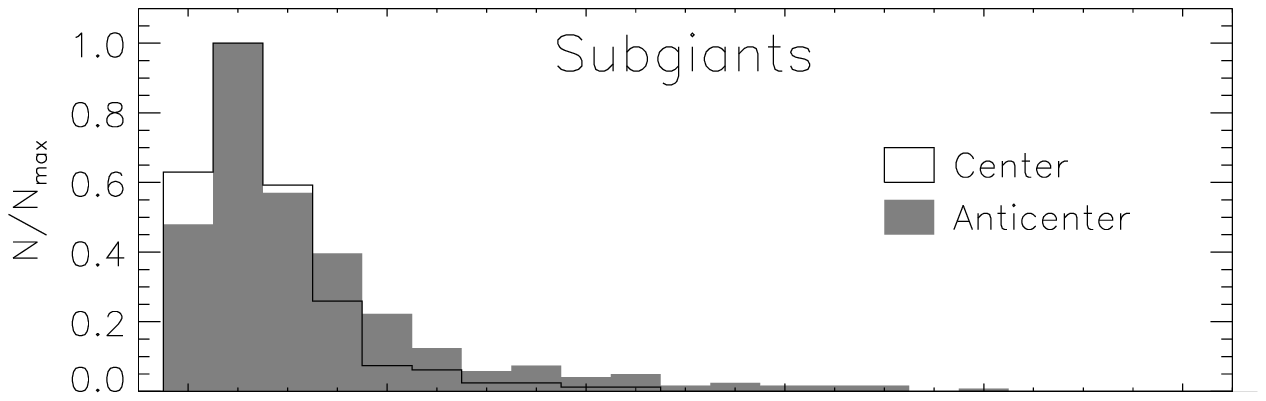}
\includegraphics[width=\columnwidth,height=3.8cm]{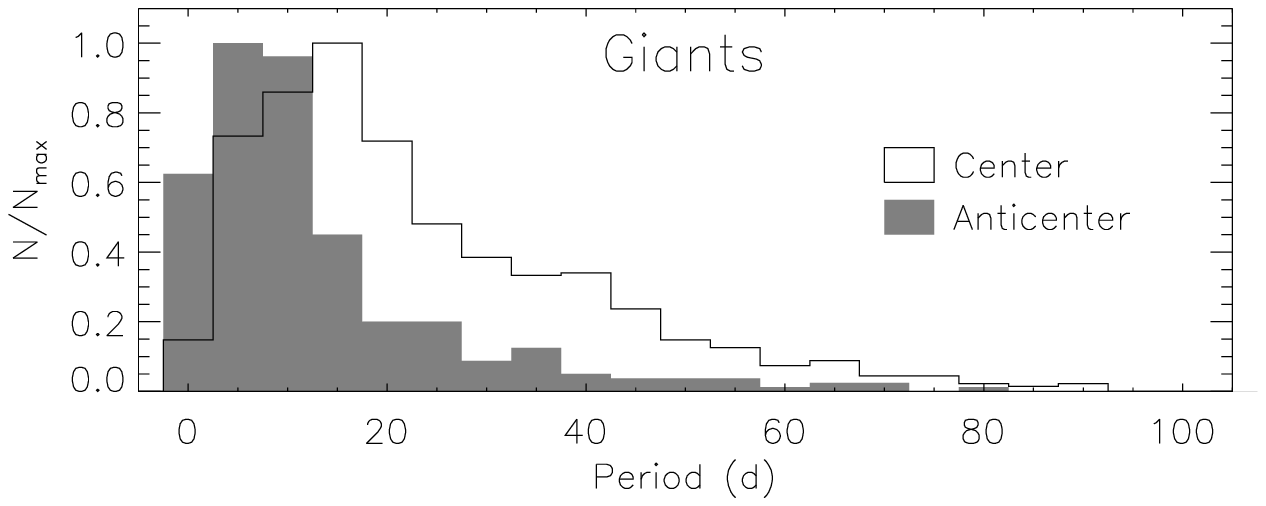}
\caption{
Distributions of the variability period in the sample described in Sect.~\ref{secobs}.
Distributions are normalized with respect to their maxima and are compared for the Galactic center and anti-center
in accordance with the symbols in the legend.
From top to bottom: full sample, main-sequence stars, subgiants, and giants.
}
\label{figcenant}
\end{center}
\end{figure}

\begin{figure}
\begin{center}
\includegraphics[width=7.9cm]{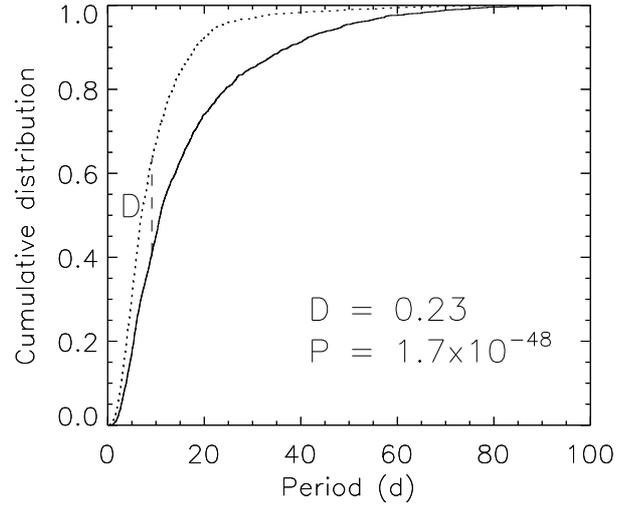}
\caption{
Cumulative distributions of the variability period for the entire sample analyzed in the top panel of Fig.~\ref{figcenant}.
Distributions are compared for Galactic center (solid line) and anti-center (dotted line)
using the Kolmogorov--Smirnov (KS) test.
Distance (D) and probability (P) calculated in the KS test are shown in the figure.
}
\label{figks}
\end{center}
\end{figure}

Here we obtain some statistics on the distribution of the
computed variability period. Fig.~\ref{figcenant} shows the
period distribution for the entire sample of 4,206 and for different
luminosity classes, with stars segregated according to Galactic region,
namely Galactic center and anti--center. The upper panel in Fig.~\ref{figcenant},
where the three luminosity classes V, IV, and III are
combined, shows that both distributions peak at around ten days and
decrease rapidly for increasing periods. Nevertheless,
for stars located in the Galactic center there is an excess of long
periods compared with those in the Galactic anti--center.
To determine whether the present data sets for the Galactic center and
anti--center are significantly different, we
performed a Kolmogorov--Smirnov (KS) test (Press et al.~\cite{pre92}), which
calculates the probability that two distributions are
derived from the same parent distribution. Fig.~\ref{figks} shows the
cumulative functions for the two variability period distributions.
The probability value of $1.7 \times 10^{-48}$ obtained by the KS test indicates that
the two distributions are in fact not drawn from the same
population distribution function. In addition, KS
analyses were conducted by comparing stars in the Galactic center and anti--center
according to luminosity class. Probability values shown in Fig.~\ref{figks}
indicate a scenario where the variability period distributions for stars in the referred Galactic regions, when compared by
luminosity class, are in fact not derived from the same parent distribution.
This result reinforces the scenario observed in Fig.~\ref{figcenant}, with a clear
excess of long periods among stars located in the Galactic center
compared with those in the anti--center.
Of course one could question whether the difference in these distributions
is produced by biases related to the LC time spans.
Although long and short runs are found either in the Galactic center or in the anti-center direction,
there are several long runs in the center direction
with shorter time spans than usual (LRc03--06; see table~\ref{tab_sample}).
This could limit the sample to shorter periods in that region,
but longer periods are found in the center direction.
Therefore, the difference in the period distributions
does not seem to be caused by biases and
may have a physical explanation with a similar discussion as in Sect.~\ref{subvarbeh}.
The explanation is possibly related to the fact that more population II stars lie
in the Galactic center than in the anti-center direction.

\section{Conclusions and future work}
\label{conclusions}

This study presents an overview of stellar LCs obtained by CoRoT within a wide range of period, color, and variability amplitude.
This is the first time that a homogeneous set of stellar variability measurements, obtained using only one instrument,
has been analyzed for a large sample with wide ranges of period, variability amplitude, and color, taking into account the 
effects of reddening on the results. As such, we were able to demonstrate the global distribution of these parameters in a representation valid for field stars.

A total of 124,471 LCs were analyzed, from which we selected a sample of 4,206 LCs presenting well-defined semi-sinusoidal signatures. Each LC was treated individually by correcting trends, outliers, and discontinuities. Through Lomb-Scargle periodograms, harmonic fits, and visual inspection, we selected the most likely periods for each variability.
Our sample shows periods ranging from $\sim$0.33 to $\sim$92 days, and variability amplitudes between $\sim$0.001 and $\sim$0.5 magnitudes, for FGKM stars with (J~--~H) from $\sim$ 0.0 to 1.4.

The color-period diagrams of the sample indicate several aspects compatible with rotational modulation. The increase in variability amplitude around (J~--~H)~$\simeq$~0.55 corroborates studies on rotating variable stars by Eker et al.~(\cite{eke08}) and  Gilliland et al.~(\cite{gil09}). The overall behavior of the increasing period with rising color index is compatible with theoretical predictions of stellar rotation. Results from this investigation were compared with public data for field variables by Hartman et al.~(\cite{har10}). The distribution periods and variability amplitudes reported here are compatible with data in the corresponding color range.
In addition to our overall results, we identified a subset of three Sun--like candidates in the context of photometric period and color, which may be of particular interest for future studies.
Moreover, we analyzed a subsample of more than 400 M-type giant stars, whose behavior seems compatible with recent studies of rotational modulation.
In addition, the distribution of variability periods for the CoRoT targets tends to be different when compared
with Galactic center and anti-center directions.
Finally, the behavior of the variability period distribution in the period--color diagram appears to substantially depend on reddening correction, which
may significantly affect age--period analyses such as that carried out in Affer et al.~(\cite{aff12}).

Observations of apparently bright stars generally provide information concerning intrinsically bright stars.
The CoRoT mission makes the important contribution of increasing the sample of intrinsically faint stars and accumulates a large amount of micro-variability data for the sources.
This demonstrates the importance of this work for studying the general variability for a significant sample of intrinsically faint field stars.
Moreover, this investigation enables future studies of the particular case of stellar rotation.

Although in many respects our results match those expected for rotating stars, photometric data alone are insufficient for identifying the physical nature of the variabilities. Therefore, additional research is necessary to confirm the root--cause of the variabilities.
As part of future research, we will combine our database with a set of spectroscopic observations currently under analysis by our team. This will allow a more accurate assessment of the results, particularly with regard to stellar rotation.

\begin{acknowledgements}

CoRoT research activities at the Federal University of Rio Grande do 
Norte are supported by continuous grants of CNPq and FAPERN Brazilian
agencies and by the INCT-INEspa\c{c}o. I. C. L. acknowledges a Post-Doctoral 
fellowship of the CNPq; C. E. F. Lopes and S. V. acknowledge 
graduate fellowships of CNPq; C. Cort\'es, J. P. B., S. C. M. and D. B. F. 
acknowledge graduate fellowships of CAPES agency; E. J.-P. and A. V. 
acknowledge financial support of the FAPESP agency. 
G.F.P.M. acknowledges the financial support by CNPq (476909/2006-6
and 474972/2009-7) and FAPERJ (APQ1/26/170.687/2004) grants.
M.C. and C.E.F.L. acknowledge support by the Chilean Ministry for the
Economy, Development, and Tourism's Programa Iniciativa Cient\'{i}fica
Milenio through grant P07-021-F, awarded to The Milky Way Millennium
Nucleus; by the BASAL Center for Astrophysics and Associated Technologies
(PFB-06); by Proyecto Fondecyt Regular \#1110326; and by Proyecto Anillo
ACT-86.
The authors warmly thank the CoRoT Technical and Manager Staffs for the development, 
operation, maintenance and success of the mission. This work 
used the SIMBAD Astronomical Database operated at the CDS, Strasbourg, France.

\end{acknowledgements}


\begin{thebibliography}{}
\bibitem[2012]{aff12} Affer, L., Micela, G., Favata, F. et al. 2012, MNRAS, 424, 11
\bibitem[2011]{bas11} Basri, G., Walkowicz, L. M., Batalha, N., et al. 2011, AJ, 141, 20
\bibitem[2006]{cas06} Casagrande, L., Portinari, L., Flynn, C., 2006, MNRAS, 373, 13
\bibitem[2011]{cat11} Catelan, M., Minniti, D., Lucas, P. W. et al. 2011 in Carnegie Observatories Astrophysics Series, 5, 145
\bibitem[2007]{deb07} Debosscher, J., Sarro, L. M., Aerts, C., et al. 2007, A\&A, 475, 1159
\bibitem[2009]{deb09} Debosscher, J.,L. Sarro, M., LÛpez, M., et al. 2009, A\&A, 506, 519
\bibitem[2009]{deg09} Degroote, P., Aerts, C., Ollivier, M., et al. 2009, A\&A, 506, 471
\bibitem[1983]{dwo83} Dworetsky, M.M. 1983, MNRAS 203, 917
\bibitem[2008]{eke08} Eker, K., Filiz-Ak, N., Bilir, S. et al. 2008, MNRAS, 389, 1722 
\bibitem[2012]{eks12} Ekstr\"om, S., Georgy, C., Eggenberger, P. et al. 2012, A\&A, 537, A146
\bibitem[2010]{gaz10} Gazzano, J.-C., de Laverny, P., Deleuil, M. et al. 2010, A\&A, 523, A91
\bibitem[2009]{gil09} Gilliland, R. L. 2009, AJ, 136, 566
\bibitem[2011]{gug11} Guggenberger, E., Kolenberg, K., Chapellier, E. et al. 2011, MNRAS, 415, 1577
\bibitem[2010]{har10} Hartman, J. D., Bakos, G. A., Kovacs, G. et al. 2010, MNRAS 408, 475
\bibitem[2009]{har09} Hartman, J. D., Gaudi, B. S., Pinsonneault, M. H., et al. 2009, ApJ 691,342
\bibitem[2003]{her03} Herwig, F.,  Langer, N., Lugaro, M. 2003, AJ, 593, 1056
\bibitem[2006]{hol06} Holmberg, J., Flynn, C., Portinari, L., 2006, MNRAS, 367, 449
\bibitem[2001]{hun01} H\"unsch, M. 2001, in Astron. Ges. Abstr. Ser., 18, MS 07 10
\bibitem[2011]{irw11} Irwin, J., Berta, Z. K., Burke, C. J. et al. 2011, ApJ, 727, 56
\bibitem[2009]{lan09} Lanza, A. F., Aigrain, S., Messina, S. et al. 2009 A\&A 506, 255
\bibitem[2010]{lan10} Lanza, A. F., Bonomo, A. S., Moutou, C. et al. 2010, A\&A, 520, A53
\bibitem[2011]{lan11} Lanza, A. F., Bonomo, A. S., Pagano, I. et al. 2011, A\&A 525, A14
\bibitem[2007]{lan07} Lanza, A. F., Bonomo, A. S., \& Rodon\`o, M. 2007, A\&A, 464, 741
\bibitem[2003]{lan03} Lanza, A. F., Rodon\`o, M., Pagano, I. et al. 2003, A\&A, 403, 1135
\bibitem[1944]{lev44} Levenberg, K. 1944, Quarterly of Applied Mathematics, 2, 164
\bibitem[1976]{lom76} Lomb, N. R. 1976 Ap\&SS, 39, 447
\bibitem[1963]{mar63} Marquardt, D. W. 1963, SIAM Journal of Applied Mathematics, 11, 431
\bibitem[2011]{mei11} Meibom, S., Barnes, S. A., Latham, D. W. et al. 2011, ApJ, 733, L9
\bibitem[2009]{mei09} Meibom, S., Mathieu, R. D., \& Stassun, K. G. 2009, ApJ, 695, 679
\bibitem[2010]{mis10} Mislis, D., Schmitt, J. H. M. M., Carone, L. et al. 2010, A\&A, 522, A86
\bibitem[2004]{pie04} Pietrinferni, A., Cassisi, S. Salaris, M. et al. 2004 ApJ, 612, 168
\bibitem[2009]{pre92} Press, W. H., Teukolsky, S. A., Vetterling, W. T. et al. 1992, in Numerical recipes, Cambridge University Press
\bibitem[2008]{ren08} Renner, S., Rauer, H., Erikson, A. et al. 2008, A\&A 492, 617
\bibitem[2008]{rie08} Rieke, G., Blaylock, M., Decin, L., et al, 2008, AJ, 135, 2245
\bibitem[2007]{sam07} Samadi, R., Fialho, F., Costa, J. E. S. et al. 2007, arXiv:astro-ph/0703354
\bibitem[2009]{sar09} Sarro, L. M., Debosscher, J., L\'opez, M. et al. 2009, A\&A, 494, 739
\bibitem[1982]{sca82} Scargle, J. D. 1982, ApJ, 263, 835
\bibitem[2011]{sil11} Silva-Valio, A. \& Lanza, A. F. 2011 A\&A 529, A36
\bibitem[2009]{str09} Strassmeier, K. G. Astron Astrophys Rev 2009, 17, 251
\bibitem[2010]{zhe10} ZhengShi, Z., YuQiin, C., JingKun, Z. et al. 2010, SciChina, 53, 579
\bibitem[2012]{val12} Valcarce, A. A. R., Catelan, M., Sweigart, A. V. 2012 A\&A, 547, A5


\end{thebibliography}
\end{document}